\documentclass[twocolumn]{aastex631}
\usepackage{gensymb}

\usepackage{amsmath}
\usepackage{threeparttable}
\newcommand{\kms}{km\,s$^{-1}$}

\begin{document}

\title{Discovery of a proto-white dwarf with a massive unseen companion}

\correspondingauthor{Gautham Adamane Pallathadka}
\email{gadaman1@jh.edu}

\author[0000-0002-5864-1332]{Gautham Adamane Pallathadka}
\affiliation{William H. Miller III Department of
Physics \& Astronomy, Johns Hopkins University, 3400 N Charles St, Baltimore, MD 21218, USA}

\author[0000-0002-0572-8012]{Vedant Chandra}
\affiliation{Center for Astrophysics $\mid$ Harvard \& Smithsonian, 60 Garden St, Cambridge, MA 02138, USA}

\author[0000-0001-6100-6869]{Nadia L. Zakamska}
\affiliation{William H. Miller III Department of
Physics \& Astronomy, Johns Hopkins University, 3400 N Charles St, Baltimore, MD 21218, USA}

\author[0000-0003-4250-4437]{Hsiang-Chih Hwang}
\affiliation{School of Natural Sciences, Institute for Advanced Study, Princeton, 1 Einstein Drive, NJ 08540, USA}

\author[0000-0002-0632-8897]{Yossef Zenati}
\affiliation{William H. Miller III Department of
Physics \& Astronomy, Johns Hopkins University, 3400 N Charles St, Baltimore, MD 21218, USA}

\author[0000-0001-5941-2286]{J. J. Hermes}
\affiliation{Department of Astronomy, Boston University, 725 Commonwealth Ave., Boston, MA 02215, USA}

\author[0000-0002-6871-1752]{Kareem El-Badry}
\affiliation{Department of Astronomy, California Institute of Technology, 200 E. California Blvd., Pasadena, CA 91125, USA}

\author[0000-0002-2761-3005]{Boris T. G\"{a}nsicke}
\affiliation{Department of Physics, University of Warwick, Coventry CV4 7AL, UK}

\author[0000-0002-6770-2627]{Sean Morrison}
\affiliation{Department of Astronomy, University of Illinois at Urbana-Champaign, Urbana, IL 61801, USA }

\author[0000-0002-8866-4797]{Nicole R. Crumpler}
\affiliation{William H. Miller III Department of
Physics \& Astronomy, Johns Hopkins University, 3400 N Charles St, Baltimore, MD 21218, USA}

\author[0000-0002-6270-8624]{Stefan Arseneau}
\affiliation{William H. Miller III Department of
Physics \& Astronomy, Johns Hopkins University, 3400 N Charles St, Baltimore, MD 21218, USA}


\begin{abstract}
\noindent
We report the discovery of SDSS~J022932.28+713002.7, a nascent extremely low-mass (ELM) white dwarf (WD) orbiting a massive ($> 1\,M_\odot$ at 2$\sigma$ confidence) companion with a period of 36 hours. 
We use a combination of spectroscopy, including data from the ongoing SDSS-V survey, and photometry to measure the stellar parameters for the primary pre-ELM white dwarf. 
The lightcurve of the primary WD exhibits ellipsoidal variation, which we combine with radial velocity data and $\tt{PHOEBE}$ binary simulations to estimate the mass of the invisible companion. 
We find that the primary WD has mass $M_1$ = $0.18^{+0.02}_{-0.02}$ M$_\odot$ and the unseen secondary has mass $M_2$ = $1.19^{+0.21}_{-0.14}$ M$_\odot$. 
The mass of the companion suggests that it is most likely a near-Chandrasekhar mass white dwarf or a neutron star. It is likely that the system recently went through a Roche lobe overflow from the visible primary onto the invisible secondary. The dynamical configuration of the binary is consistent with the theoretical evolutionary tracks for such objects, and the primary is currently in its contraction phase. The measured orbital period puts this system on a stable evolutionary path which, within a few Gyrs, will lead to a contracted ELM white dwarf orbiting a massive compact companion. 

\end{abstract}



\section{Introduction}
White dwarfs (WDs) are remnants left behind by main sequence stars with masses in the range 0.8-8 M$_{\odot}$. While the typical masses of white dwarfs range from 0.5 M$_{\odot}$ up to 1.4 M$_{\odot}$, extremely low mass WDs (ELM WDs) with masses in $0.15-0.35$ M$_{\odot}$ have been detected in binaries \citep{Iben_1985,Brown_2010,Istrate_2016,el-badry_birth_2021}. These ELM WDs are usually formed when a massive main sequence star loses its outer envelope to a companion. The helium core that remains with a hydrogen atmosphere is an ELM WD \citep{Nelson_2004,li_formation_2019}. They have been theorized to form only in binary systems --- their mass is too low to be formed from single star evolution within a Hubble time --- and this scenario is supported by observations as well \citep{Marsh_1995,Brown_2010}. 

ELM WDs can be found in binaries with WDs \citep{Brown_2020}, pulsars \citep{Nelson_2004,Antoniadis_2013,istrate_formation_2014} or main sequence stars \citep{Maxted_2011}. These binaries are interesting because they help us piece together the evolutionary pathways and test our understanding of binary evolution. Compact binaries of double WDs are particularly interesting since they are possible Type Ia supernova progenitors \citep{Nomoto_1984,maoz_observational_2014,Soker_2019,Jha_2019}. They are also dominant sources of low-frequency gravitational waves for future observatories like the Laser Interferometer Space Antenna (LISA; \citealt{Marsh_2011,Korol_2022}). WD mergers with neutron stars (NS) or black holes are candidate progenitors of gamma ray bursts or other types of explosive transients \citep{fryer_merging_1999,Margalit_2016,Bobrick_2017,Zenati_2019,Zenati_2020,Bobrick_2022,Kaltenbourn_2022}. In binaries with pulsars, the mass transfer phase can lead to spin-up of the pulsar giving birth to milli-second pulsars. These systems can be used to independently verify the spin-down age of milli-second pulsars \citep{Nelson_2004}. 

While ELM WDs have surface gravity values in the range $\log g=5 - 6.5$ (where $g$ is in cgs units), in the early stages of their evolution they can appear as bloated pre-ELM WDs with a smaller $\log g$ \citep{Lagos_2020, Wang_2020, Kupfer_2020a,Kupfer_2020b,el-badry_birth_2021}. Using spectroscopic and photometric data one can solve for the parameters of the visible object, hereafter called the primary. Without additional information it is not possible to solve for the mass of the invisible -- secondary -- object because of a degeneracy between the mass of the secondary and the system's inclination. Because of their bloated nature, the surfaces of pre-ELM WDs can be tidally distorted in the presence of a companion. This change in the shape of stellar surface leads to a periodic variation in the lightcurve. Therefore, the mass-inclination degeneracy can be further constrained with light curve data. 

In this paper we report the discovery of SDSS~J022932.29+713002.6 (hereafter SDSS~J0229+7130), a binary consisting of a bloated pre-ELM WD and an invisible companion with a mass close to and perhaps exceeding the Chandrasekhar mass. We summarize the observations in Sec. \ref{sec:obs}. We analyze them and present the measurements of the binary parameters in Sec. \ref{sec:analysis}. We discuss the results in Sec. \ref{sec:discussion}. Throughout the paper $g$ is the acceleration due to gravity in cgs units (cm s$^{-2}$), [Fe/H] is the abundance ratio relative to the Sun on a logarithmic scale, $Z$ is the metallicity of the star, i.e. the mass fraction of metals on a linear scale, and $\mathrm{T_{eff}}$ is the stellar temperature in Kelvin. Inclination angles are measured relative to the plane of the sky, so that $i=90\deg$ is an edge-on system. We refer to the less massive WD as being the primary component, and the secondary is the more massive invisible companion.

\section{Observations and data reduction}
\label{sec:obs}

\subsection{Spectroscopic Data}
We identified SDSS~J0229+7130 in the first-year data from the Milky Way Mapper, a multi-epoch Galactic spectroscopic program in the fifth-generation Sloan Digital Sky Survey \citep[SDSS-V,][]{kollmeier_sdss-v_2017,Kollmeier_2019}. Milky Way Mapper started in November 2020 using the Apache Point Observatory 2.5 m telescope \citep{gunn_2006} and is now also operating at the Las Campanas Observatory 2.5 m telescope \citep{bowen_1973} using both the Baryon Oscillation Spectroscopic Survey spectrograph (BOSS; \citealt{Smee_2013}) and the Apache Point Observatory Galactic Evolution Experiment (APOGEE) spectrographs \citep{wilson_2019}. Our target was observed for a total of 9 individual exposures between 2020-12-04 to 2020-12-06 with BOSS. Each exposure lasted 900s and covered a wavelength range from 3600 \AA\ to 10000 \AA\ at a resolution of $R\simeq 1800$. The wavelengths of the spectra are corrected to the heliocentric frame, and the absolute wavelength calibration of each exposure is accurate to $<$10 km s$^{-1}$. The BOSS data products in this paper were derived by using \verb|IDLspec2D v6_1_0|.

\begin{figure*}[ht!]
    \centering
    \includegraphics[width=\linewidth]{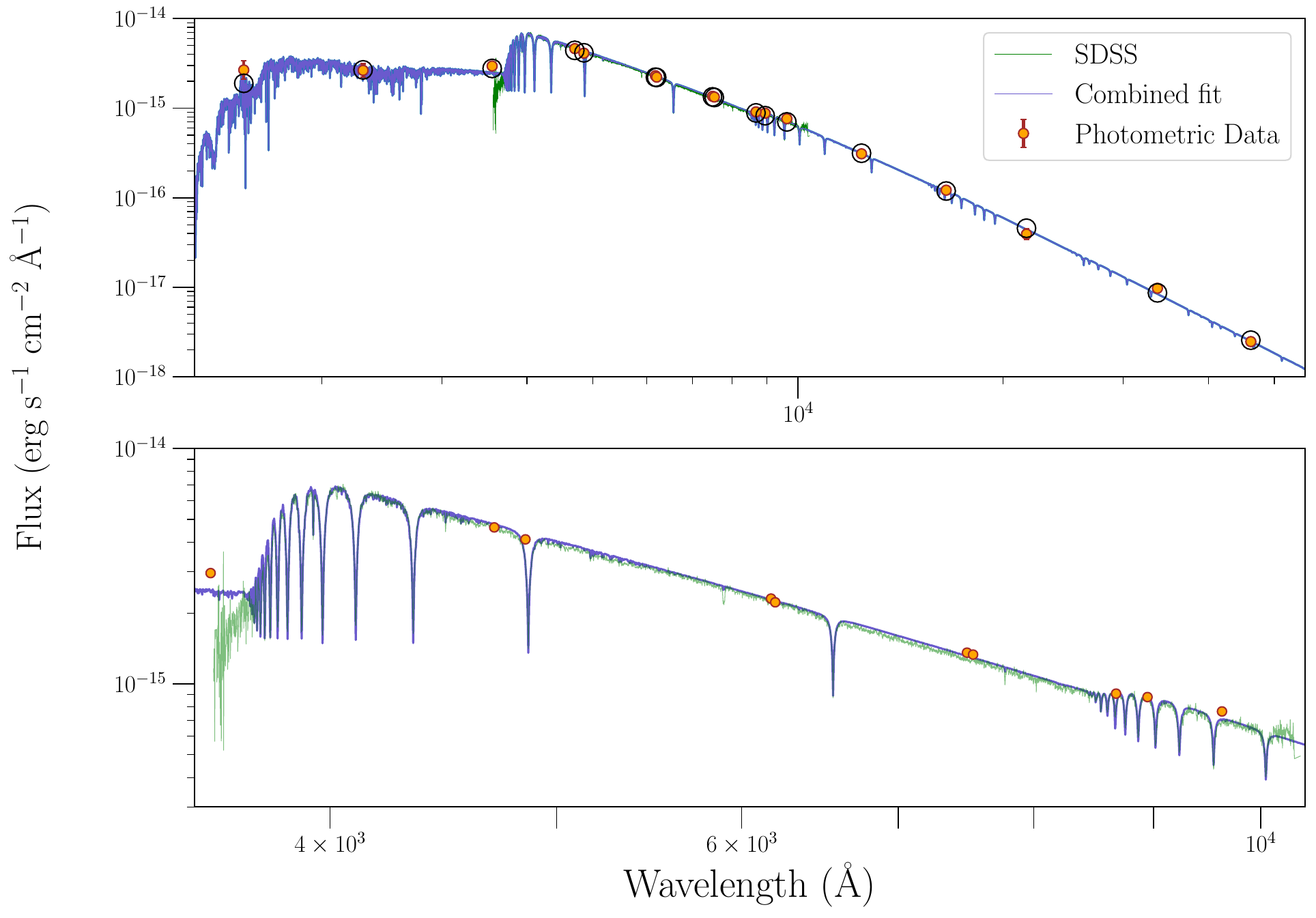}
    \caption{The flux-corrected coadded SDSS spectrum (green line) and archival photometry (orange points) are shown along with the best-fit theoretical spectrum (blue line). The bottom panel shows the spectral region covered by the SDSS spectroscopic data.}
    \label{fig:bosz_fit}
\end{figure*}

We initially flagged SDSS~J0229+7130 due to its significant radial velocity (RV) variation $\gtrsim 100$~\kms{} between different nights, albeit with minimal variation across successive exposures on a given night. This suggested that both the RV amplitude and orbital period are large, hallmarks of a high mass function binary. Spectroscopically, SDSS~J0229+7130 looks like a pure hydrogen atmospheric DA White Dwarf, with narrow Balmer lines (Fig. \ref{fig:bosz_fit}). In addition, there is a faint sodium absorption doublet \ion{Na}{1} D near 6000 \AA\ and calcium absorption line \ion{Ca}{2} K near 3933 \AA, which are discussed further in Section \ref{subsec:Stellar_Param}. 
\begin{figure}
    \centering
    \includegraphics[width=\columnwidth]{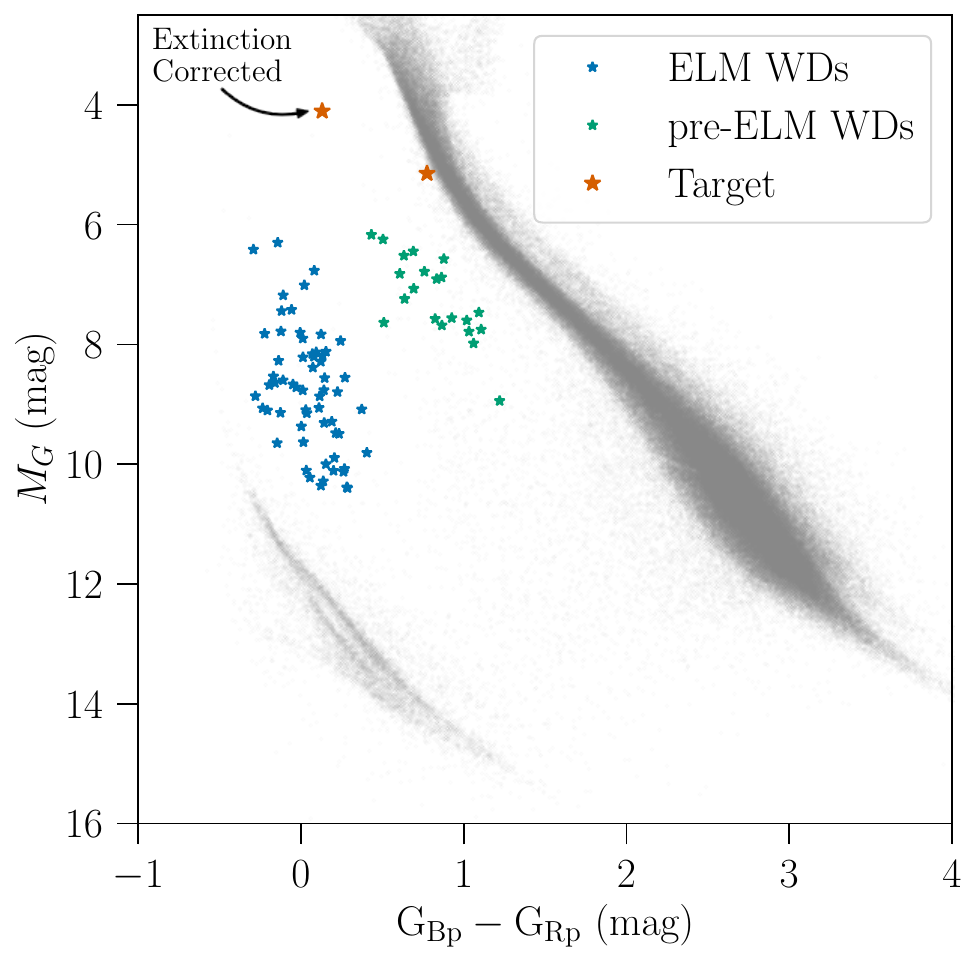}
    \caption{Color-magnitude diagram. Our target is marked in brown. We also show the extinction-corrected position. The blue points are ELM WDs from \cite{Brown_2020} and the green points are the pre-ELM WDs from \cite{el-badry_birth_2021}. The sequence in the lower left, below the ELM candidates, is the normal WD track. Our target is just below the main sequence.}
    \label{fig:color-mag}
\end{figure}

Although the SDSS-V data suggested a high mass-function binary, the RV data were too sparse to better constrain the mass function and definitively measure the orbital period. To fill in the orbital phase coverage, we obtained follow-up spectroscopy with the Dual-Imaging Spectrograph (DIS) on the 3.5 m telescope at the Apache Point Observatory (APO). We used B1200/R1200 gratings with a $1.5''$ slit, delivering resolution $R \approx 3000$. We obtained twenty 900s exposures between three different nights on 2021-10-05, 2021-11-01 and 2021-11-05. The APO data were reduced by the standard IRAF pipeline. 

The absolute wavelength calibration of the APO spectra showed systematic shifts, which increased with the wavelength. To account for this, we estimate the shift at 6498 \AA\ (close to the H$\alpha$ line) using the known positions of the sky lines, and measure the RVs incorporating this derived correction. We also increase the RV error by adding this correction in quadrature to the measured RV error, to get the most conservative RV uncertainties. The RVs for APO data are corrected to barycentric frame using the package \texttt{barycorrpy}\footnote{\url{https://github.com/shbhuk/barycorrpy}} \citep{lia_corrales_2015_15991, Kanodia_2018}. The low wavelength end of APO spectrum had issues and is unreliable. We only use the data with wavelength greater than 5900 \AA.

\subsection{Photometric Data}
Our target is at a distance of $d=1.66 \pm 0.12$ kpc, which is calculated by \citet{Bailer-Jones_2021} using the {\it Gaia} parallax measurement \citep{Gaia_DR3}. Looking at the {\it Gaia} color-magnitude diagram (Fig. \ref{fig:color-mag}), we see that the object lies well above the white dwarf track, and is only slightly below the main sequence. Thus, based on the spectrum and the position of the star on the \textit{Gaia} color-magnitude diagram, we propose that this object is neither a typical WD nor a main sequence star. There are no resolved wide companions (at separation $>$2000 AU) near our target in {\it Gaia}, and there are no astrometric anomalies reported in {\it Gaia} DR3 for this source. So there is no evidence for additional companions to the spatially unresolved system. The object is close to the plane of the Galaxy with b $\approx$ 10\degree, and given the inferred distance we need to carefully take into account the interstellar extinction, as explained in detail in Section \ref{sec:analysis}.

In addition to the spectroscopic data, this object has archival Zwicky Transient Facility (ZTF) observations \citep{masci_zwicky_2019,ZTF_doi}, and we selected all observations released in DR18. The object was observed with the $g$ filter between 2018-06-29 and 2023-02-20, and between 2018-06-14 and 2023-03-10 with the $r$ filter. To select clean data, we choose all point sources in ZTF within 5" of target position obtained from {\it Gaia} with ZTF quality flag $\tt{catflags}$ = 0 and $\vert \tt{sharp} \vert <$  0.25. We find that all the available data points correspond to our object and are within 5" of our target. However, quality cuts reduce the number of data points from 912 to 821, and we obtain a clean ZTF lightcurve. 

This object also has archival Transiting Exoplanet Survey Satellite (TESS; \citealt{TESS_2015}) observations in the Full Frame Image (FFI) data in sectors 18 and 19 \citep{TESS_18_DOI,TESS_19_DOI}. The data in sector 18 have instrumental artifacts and hence we restrict ourselves to data from sector 19. We obtain the lightcurve from FFI using $\tt{eleanor}$\footnote{\url{http://adina.feinste.in/eleanor/}} \citep{eleanor_2019,astrocut_2019}. 

We collect archival photometric data for this object using VizieR\footnote{\url{http://vizier.u-strasbg.fr/cgi-bin/VizieR}} \citep{Ochsenbein_2000}. This object has photometric data from the Sloan Digital Sky Survey in $ugriz$ filters \citep{SDSS_DR16}, from 2MASS in $J, H, K_s$ filters \citep{2MASS_2006}, from GALEX in FUV and NUV bands \citep{GALEX_2017}, from Pan-STARRS in $grizy$ filters \citep{PanSTARRS_2018} and from CatWISE in W1, W2 bands \citep{Wright_2010, CatWISE_2020} covering the UV to infrared. These data are summarized in Table~\ref{tab:phot_dat}. We add a base error of 0.03 mags in quadrature to the existing errors to take into account systematics.
\begin{table}
    \caption{Spectral Energy Distribution of SDSS~J0229+7130}
    \centering
    \begin{threeparttable}
    \begin{tabular}{c c} 
         \hline \hline
         Filter & AB Magnitude \\ \hline 
         GALEX FUV & 21.36 $\pm$ 0.25\\
         GALEX NUV & 20.80 $\pm$ 0.18\\
         SDSS u & 18.15 $\pm$ 0.03\\
         SDSS g & 16.62 $\pm$ 0.03\\
         SDSS r & 16.29 $\pm$ 0.03\\
         SDSS i & 16.17 $\pm$ 0.03\\
         SDSS z &  16.06 $\pm$ 0.03\\
         Pan-STARRS g & 16.62 $\pm$ 0.03\\
         Pan-STARRS r & 16.32 $\pm$ 0.03\\
         Pan-STARRS i & 16.17 $\pm$ 0.03\\
         Pan-STARRS z & 16.12 $\pm$ 0.03\\
         Pan-STARRS y & 15.98 $\pm$ 0.03\\
         2MASS J  &  16.23 $\pm$ 0.05\\
         2MASS H &  16.50 $\pm$ 0.11\\
         2MASS Ks &  17.07 $\pm$ 0.16\\
         Cat-WISE W1 & 17.56 $\pm$ 0.03\\
         Cat-WISE W2 & 18.34 $\pm$ 0.03\\
    \hline \hline 
    \end{tabular}
    \begin{tablenotes}
     \item[*] Conversion from Vega to AB magnitude system is performed by using results from \cite{WISE_2011} for WISE and \cite{Blanton_2007} for 2MASS
   \end{tablenotes}
    \end{threeparttable}
    \label{tab:phot_dat}
\end{table}

\section{Analysis}
\label{sec:analysis}

In this section we describe the steps followed in deriving the stellar parameters of the primary using spectroscopic and photometric data. Following this, we combine the spectroscopic analysis with the analysis of the ellipsoidal variation from the photometric data to obtain the mass of the secondary. 

\subsection{Stellar Parameters}
\label{subsec:Stellar_Param}

The SDSS spectra for the target show periodic shifts in the position of absorption lines suggesting a varying radial velocity (RV). We choose to measure RV using H${\alpha}$ and H${\beta}$ simultaneously from SDSS-V spectra. We use just the H${\alpha}$ for APO spectra because the spectrum at low wavelength end is unreliable. We use \texttt{corv}\footnote{\url{https://github.com/vedantchandra/corv}} to measure the RVs for our SDSS and APO sub-exposures. \texttt{corv} computes the cross-correlation between the observed stellar spectra and the template.  For the template, we model each absorption line using two Voigt profiles with a common centroid. Once we obtain the RVs, we correct each spectrum for the Doppler shift associated with the measured RV and then coadd them in the target's rest-frame. One of the SDSS sub-exposures has flux calibration issues and we do not include it in coadding, but we still use it for the RV measurements since the relative flux calibration does not affect the position of the absorption lines. The coadded SDSS spectrum is shown in Fig.~\ref{fig:bosz_fit}. 

Interestingly, the \ion{Na}{1} D doublet near 5900 \AA\ shows little sign of periodic variation, suggesting an interstellar origin rather than an association with the stellar photosphere. The measured temperature ($>$ 8000 K, Table.~\ref{tab:res}) is also far too high to observe photospheric sodium lines \citep{Yamaguchi_2023}. To verify this, we coadd RV$>$0 sub-exposures and RV$<$0 sub-exposures separately and compare the two. We confirm that the sodium absorption is stationary and is therefore associated with the interstellar medium. We then coadd all the spectra without shifting them to the rest frame of the primary and fit a Gaussian profile to the the \ion{Na}{1} D lines. We obtain equivalent widths for D1, centered at 5897.5 \AA, and D2, centered at 5891.6 \AA\, to be 0.79 \AA\ and 0.82 \AA\ respectively, and 1.62 \AA\ for the combined D1+D2. Using the best-fit empirical relation between the \ion{Na}{1} D equivalent widths and extinction from \cite{Poznanski_2012}, we derive the 2-$\sigma$ lower limit on the color excess, $E(B-V)$, to be 0.57, 0.31, 0.68 using D1, D2 and D1+D2 respectively. The trace photospheric sodium may contribute to the absorption, and we have not incorporated the inherent error in the empirical relation, so these values are only used as indicators of high extinction, which needs to be carefully taken into account in the subsequent modeling of the system. 

The \ion{Ca}{2} K line at 3933.6 \AA\ is from the stellar surface. This is a feature observed in many ELM WDs as seen in \cite{Gianninas_2014}. We find that this absorption feature shows clear RV variation.


Using 1D hydrogen-atmosphere WD model spectra\footnote{\url{http://svo2.cab.inta-csic.es/theory/newov2/index.php?models=koester2}} \citep{Koester_2010} to fit the spectrum gives a best-fitting model at the low gravity edge of the model grid, $\log g = 5$. Therefore, to measure the stellar parameters, instead of using WD models we fit the spectrum with theoretical stellar models derived for main sequence stars with lower surface gravity, as one would expect for pre-ELM objects, and with varying metallicity. To generate theoretical spectra we use the BOSZ grid \citep{bohlin_new_2017}. We use a grid with temperatures between 6000 K and 12000 K, $\log g$ between 2 and 5, [Fe/H] between -2.25 and  0.5, each with spacing of 250 K, 0.5 dex and 0.25 dex respectively. 

We start with theoretical spectra with resolution of $R\sim10000$, which are downgraded versions of theoretical spectra with $R\sim30000$ provided by \citet{bohlin_new_2017}, much greater than the SDSS spectra. We convolve the theoretical spectra with a Gaussian to match the SDSS resolution of $R \sim 2000$. Finally, we use a \texttt{scipy} function \texttt{RegularGridInterpolator} \footnote{\url{https://docs.scipy.org/doc/scipy/reference/generated/scipy.interpolate.RegularGridInterpolator.html}} \citep{Weiser_1998,2020SciPy-NMeth} to interpolate between the stellar parameter grid points. We then fit the coadded SDSS spectrum with the interpolated and convolved theoretical spectra to obtain the best-fit $\log g$, temperature and metallicity of the star. 

For the actual fitting, we minimize the $\chi^2$ between the rest-frame coadded SDSS spectrum and the theoretical spectra. The SDSS spectrum can have flux calibration issues and the extinction is not well determined. To take into account these long-wavelength features we multiply the theoretical spectrum by a polynomial function of wavelength. We examine the solution to make sure that the polynomial only corrects the long-wavelength features and does not over-fit the absorption lines. The results are presented for polynomial of order 6, the lowest order which gives a stable solution. The spectroscopic log-likelihood to maximize is given by
\begin{multline}
\log \mathcal{L_{\mathrm{spec}}}(\mathrm{T_{eff}},\log g,\mathrm{[Fe/H]}) = \\
-\frac{1}{2} \sum_{i} \frac{(f_{\mathrm{theo}}(\lambda_i) - f_{\mathrm{obs},i})^{2}}{\sigma_{\mathrm{obs},i}^2} - \frac{1}{2} \ln(2\pi\sigma_{\mathrm{obs},i}^2),
\end{multline}
where $f_{\mathrm{theo}}(\lambda_i)$ is the theoretical spectrum evaluated at different wavelengths, while $f_{\mathrm{obs},i}$ and $\sigma_{\mathrm{obs},i}$ are the spectral flux densities and the associated errors at those wavelengths.

We fit the extinction-corrected photometric magnitudes to obtain the radius, temperature and $\log g$ of the primary. We estimate the extinction using the \citet{Green_2019} model, which is a 3D dust map derived from PanSTARRS. The extinction calculation is done using \texttt{dustmaps}\footnote{\url{https://dustmaps.readthedocs.io/en/latest/}} \citep{dustmaps} and \texttt{pyextinction}\footnote{\url{https://github.com/mfouesneau/pyextinction}} packages. We infer the $(B-V)$ color excess using $E(B-V) = 0.884 \times \alpha$, where $\alpha$ is the dust reddening returned by the \texttt{dustmaps} package and represents the dust density in the line of sight \citep{Green_2018}. We then use a \cite{Fitzpatrick_1999} extinction curve with $\mathrm{R_{V}=3.1}$ to obtain the extinction in various filters. The \cite{Green_2019} dust map is probabilistic and there is a spread associated with the predicted extinction. We use the median of all returned samples of $\alpha$ by \texttt{dustmaps} to calculate the color excess. Additionally, the conversion from the $\alpha$ values from \cite{Green_2019} to $E(B-V)$ or the value of $R_V$ has uncertainty due to the varying dust size distribution and the resulting varying extinction curves. To take all of this into account we assign a conservative error of 20\% to the median value provided by \texttt{dustmaps}. We use a Gaussian prior for $E(B-V)$ with mean and sigma calculated as described above and leave it as nuisance parameter in our fitting. There are other 3D extinction maps such as \citet{Capitanio_2017} and \citet{Lallement_2019} which give somewhat different values for extinction. In particular, \cite{Lallement_2019} predicts a much larger extinction and the disagreements between different extinction maps have been reported by them. We use the predictions from \cite{Green_2019} for our fiducial solution and discuss the consequences of higher extinction in \ref{sec:discussion}.

The theoretical spectra used for the photometric fitting are generated using the BaSeL library \citep{Lejeune_1997,Lejeune_1998}, which are photometrically corrected semi-empirical models. The theoretical models estimate the flux on stellar surface. We can multiply this by a factor of $(R/d)^2$ to obtain the apparent flux, where $R$ is the radius of the primary and $d$ is the distance to the object. The computation is performed using the package \texttt{pystelllibs}\footnote{\url{https://mfouesneau.github.io/pystellibs/}}. We use the {\it Gaia} parallax and include the distance as a nuisance parameter using the prior provided by \cite{Bailer-Jones_2021}. 
We then use the \texttt{pyphot}\footnote{\url{https://mfouesneau.github.io/pyphot/}} package to integrate the theoretical spectra through the appropriate filters, and fit the extinction-corrected magnitudes with the theoretical magnitudes to estimate the best-fit parameters.  The actual fitting is again performed by likelihood maximization. We find that the photometric fits very weakly depend on metallicity, $Z$, and thus we can fix $Z = 0.014$ without affecting our results significantly. The likelihood, $
\mathcal{L}_{\mathrm{phot}}$, is defined similarly to the spectroscopic likelihood, replacing fluxes with appropriate magnitudes.

Finally, we perform joint spectrophotometric fitting by defining the combined likelihood as $\mathcal{L}(T_{\rm eff},\log g,\mathrm{[Fe/H]},R,d,E(B-V)) = \mathcal{L}_{\mathrm{phot}} \times \mathcal{L}_{\mathrm{spec}} \times \mathcal{L}_{\mathrm{priors}}$.  $\mathcal{L}_{\mathrm{priors}}$ is the likelihood associated with the priors for extinction and parallax as described earlier, and also the uniform priors for rest of the parameters with limits set by the BOSZ grid. For the fitting, we calculate log-likelihoods for each of these separately and add them. The posterior distribution is explored using \texttt{emcee}\footnote{\url{https://emcee.readthedocs.io/en/stable/}} \citep{Foreman_Mackey_2013}.

The results from fitting the spectroscopic and photometric data alone showed the existence of two local minima centered around color excess of 0.41 and 0.51 respectively, and a degeneracy between $E(B-V)$ and other stellar parameters. In Sec.~\ref{subsec:comb_fit}, we combine the spectrophotometric data with lightcurve, radial velocity and fit them simultaneously. To this end, we explore solutions around both these minima separately by restricting the color excess to the values above and setting a nominal error of 5\%.

\subsection{Lightcurve and Orbital Parameters}
\label{subsec:lc}

\begin{figure}[h]
    \centering
    \includegraphics[width=\columnwidth]{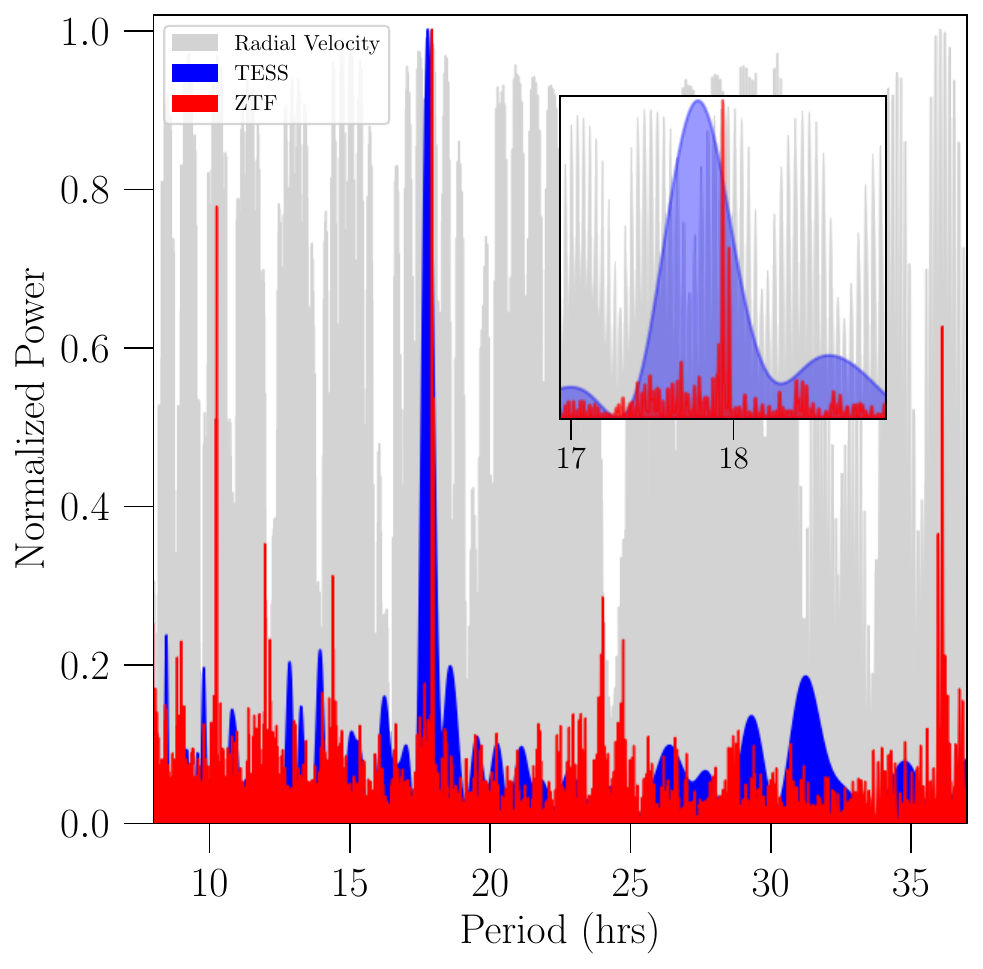}
    \caption{The normalized periodograms from various datasets: blue for TESS photometry, red for ZTF photometry, grey for spectroscopy.} 
    \label{fig:periodogram_all}
\end{figure}

We can derive the orbital parameters by analyzing the radial velocity and the lightcurve data. 
The sampling of the RVs from spectroscopy is insufficient for deriving the period, and therefore we use the lightcurve data to measure the period. Stars in close binary systems show periodic changes in flux due to eclipses, heating of the companion due to the primary (reflection binary; \citealt{Vaz_1985,Schaffenroth_2022}), or distortions of their surfaces induced by the tidal forces of the companion (ellipsoidal modulation;  \citealt{kopal_1959, green_2023}). We combine the data from both the $g$ and $r$ filters of ZTF and normalize the data to capture only the fractional variations about the median flux. We then use the Lomb-Scargle periodogram to derive the period of the orbit, and phase fold both the lightcurve as well as the RV data to this period. The periodogram is shown in Fig.~\ref{fig:periodogram_all}. The amplitude of the flux variation can differ depending on the filters used. However, for the purpose of period determination, we choose to collate data from both filters and treat them as single dataset. This results in a more robust period determination while having negligible downsides. 

\begin{figure}[h]
    \centering
    \includegraphics[width=\columnwidth]{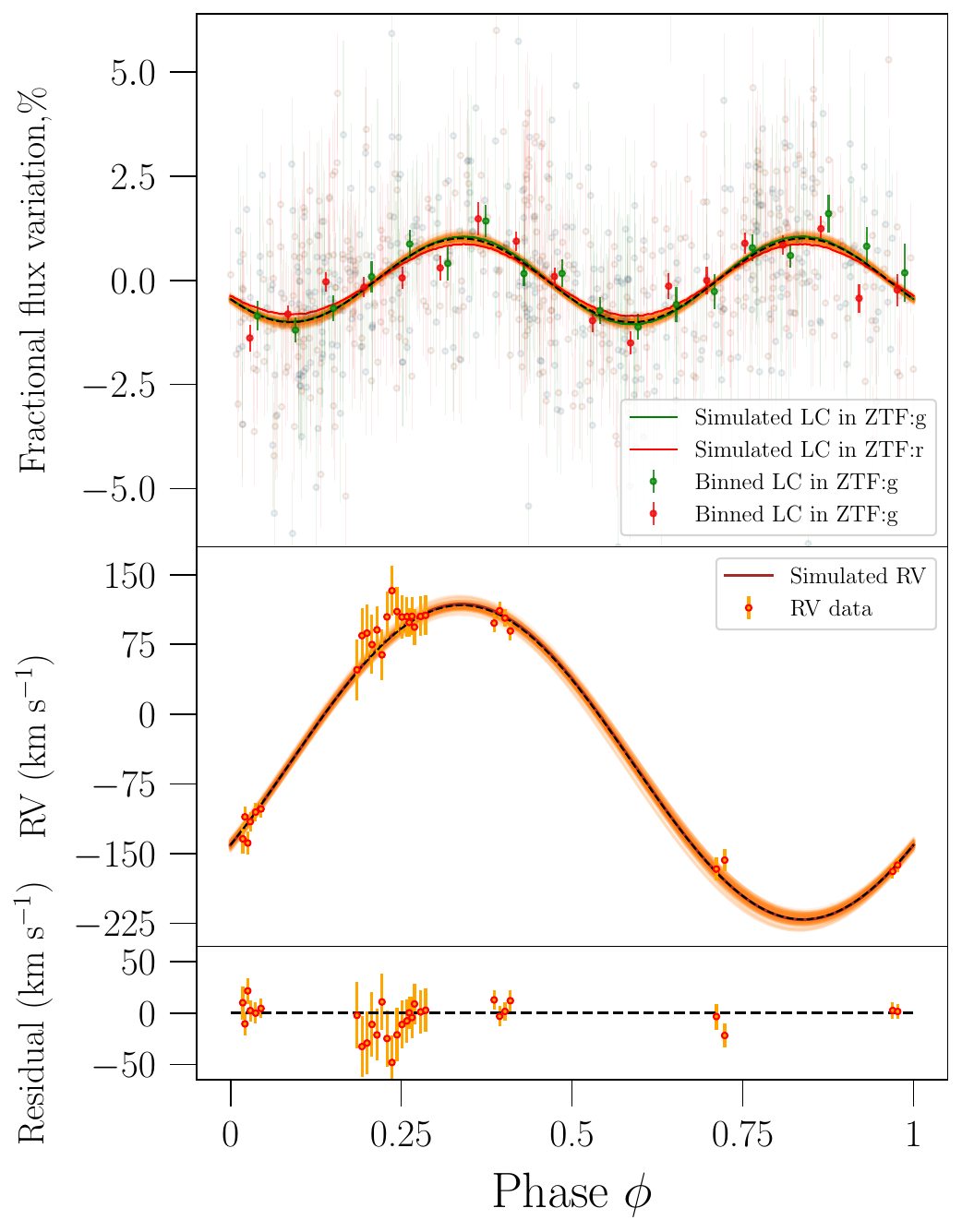}
    \caption{Radial velocity and ZTF lightcurve phase-folded at the best period. In the top plot, solid data points are the binned lightcurve while the raw data is plotted in the background. We plot the \texttt{PHOEBE} simulated RV and lightcurve data for the best-fit parameters. In the top two panels, for comparison, we also plot the best-fit sinusoidal models as black dashed lines and also plot 100 random samples of sinusoidal models, in orange, as representative of the errors involved.} 
    \label{fig:rvlc}
\end{figure}

Depending on the physical reasons that cause the flux variability, the orbital period of the system can correspond to the peak of the periodogram or be twice as long. The dominant effect of the tidal distortion is to produce a prolate ellipsoid pointed at the companion and this produces a quasi-sinusoidal variation in the emitted flux due to varying surface area. As discussed in \cite{el-badry_birth_2021}, the quasi-sinusoidal variation would have different minima at two different end-on configurations (when the longest axis of the ellipsoid is directed towards the observer) due to different gravity darkening. However, this is a second order effect which could be virtually invisible in a noisy lightcurve. If the minima are not distinguishable, then the periodogram peaks at half the orbital period. We find that phase folding to the period $P = 35.8703$ hours, twice the period corresponding to the maximum power of the periodogram, fits both the lightcurve and the RV data well, and the resulting fit is shown in Fig.~\ref{fig:rvlc}. 

The variation of both datasets behave as we would theoretically expect for ellipsoidal variation -- with one of the maxima of the lightcurve aligning with the minimum of RV curve and the other maximum coinciding with the RV maximum. This strongly supports our argument that the origin of the photometric variation is orbital in nature and is due to the ellipsoidal variation. We do not observe any signs of eclipses, which would be signaled by a difference in the depth of the minima. When we fold the data to the period corresponding to the maximum power in the periodogram, we find that the RV data are not well fit. This rules out the possibility of a reflection binary, where the photometric variability is due to the differences in the temperature of the primary and the companion \citep{Wilson_1990}. For systems where only one of the binary stars is visible, reflection effect is seen for very hot primaries ($>$ 30000K, \citealt{Hilditch_1996}). Since this is not the temperature regime we are dealing with, it is reasonable that photometric variability due to reflection is not the dominant effect in our source. 

To verify the obtained period, we repeat the process with TESS data. There are two objects at 12$^{\prime \prime}$ and 19$^{\prime \prime}$ from our target, while each TESS pixel covers a region of 21$^{\prime \prime}$. Hence, the observed flux from our source is blended with that of the contaminants and the absolute flux variability measured by TESS does not accurately represent the variability of our target. Therefore, we analyze the TESS lightcurve separately from the ZTF lightcurve to account for the potential blending with nearby sources. Indeed, we find that the fractional variability of the TESS lightcurve is  about 0.4\%, much smaller than that of the ZTF lightcurve (2\%). Thus, we restrict ourselves to using the ZTF data for further analysis while using TESS to verify the period. We obtain periods of 35.8703 $\pm$ 0.0006 hours and 35.56 $\pm$0.07 hours from ZTF and TESS data respectively. 
 
From the Lomb-Scargle periodogram, we compute the un-normalized power spectral density -- which is proportional to $\chi^2$ defined between the lightcurve data and the periodogram model \citep{Scargle_1982,VanderPlas_2018}. To estimate the errors, we calculate the frequencies where this $\chi^2$ increases by magnitude of one from the minimum. While there is a mis-match in the derived periods, at 5-$\sigma$ level they are not inconsistent. Using TESS we also verify that the period of 36 hours is robust and not associated with aliasing because of the different cadences associated with each dataset.

Once we obtain the period, to measure the amplitude of variations and to verify that the periodicity is indeed due to ellipsoidal variation, we re-fit the lightcurve and RV data with a sinusoid (i.e assuming eccentricity $e$ = 0) -- first individually and then both simultaneously. We fix the period and use the phase-folded data while leaving semi-amplitude of flux variation, RV semi-amplitude ($K$), systemic velocity ($v_{\gamma}$) and phase ($\phi$) as free parameters. We find that both variations behave exactly as we expect and are exactly in phase. We also find that the flux of our target varies by about ($2.02 \pm 0.14$)\% from the minimum to the maximum of the light curve and the RV semi-amplitude is $K =  (169 \pm 4)$ km s$^{-1}$. Once we know the RV semi-amplitude, orbital period and mass of one of the stars we can use the binary mass function to express the inclination in terms of the mass of the second star:
\begin{equation}
    i = \arcsin\bigg( \bigg(\frac{(M_1 + M_2)^2}{M_2^3} \frac{P_{orb} K^3}{2\pi G}\bigg)^{1/3} \bigg)
\end{equation}

\begin{figure}[h]
    \centering
    \includegraphics[width=\columnwidth]{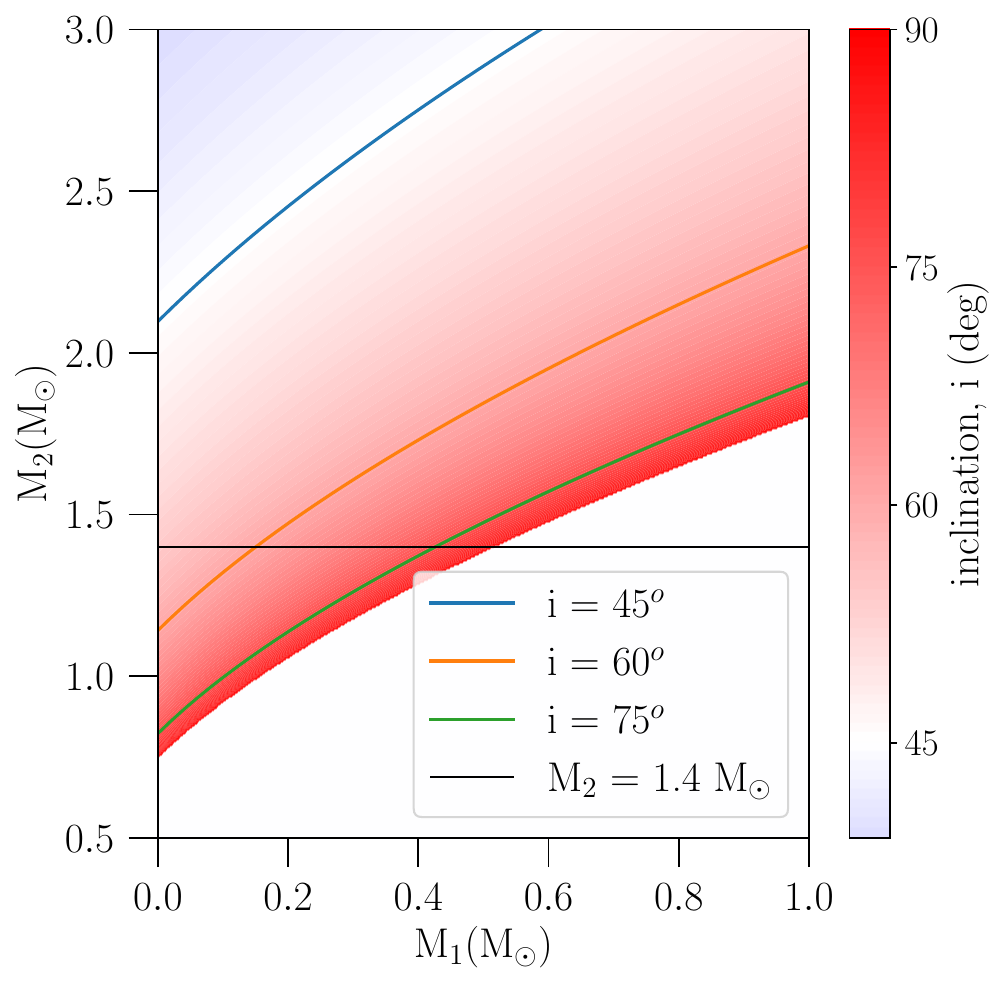}
    \caption{The parameter space of masses of the binaries allowed by the binary mass function. $M_1$ is the mass of the visible star, $M_2$ is the mass of the invisible companion, and $i$ is the orbital inclination counted from the plane of the sky. The colored regions are allowed while the white region in the bottom right would require an unphysical $\sin i >1$. We also show the Chandrashekhar limit (1.4 M$_{\odot}$) for $M_2$.}
    \label{fig:m1m2}
\end{figure}

Fig.~\ref{fig:m1m2} illustrates the parameter space that we need to explore. We see that based on the relatively long period combined with a relatively large RV variation, the secondary must be quite massive. Normally, one of the constraints on masses would be that the star should not overflow its Roche lobe radius \citep{Eggleton_1983}. However, in our case, the period of the binary is quite large and hence the orbital separation for reasonable masses (primary mass $M_1>$ 0.1 M$_{\odot}$ and secondary mass $M_2>$ 1 M$_{\odot}$) is large enough that the Roche lobe radius is much larger than the radius of the star and this constraint is never important. 

Fig.~\ref{fig:m1m2} gives a lower limit on the secondary mass $M_2$ of about 0.8 M$_{\odot}$. Given that we only see the primary, we suspect that the companion is a compact object. For any star, flux near the Rayleigh-Jeans tail of the stellar spectrum is proportional to $R^2 T$. Thus, the ratio between the flux of two stars at large wavelengths can be written as $(R_1/R_2)^2 (T_1/T_2)$. We can use this to our advantage and rule out main sequence companions. If the companion is a 0.8 $M_{\odot}$ main sequence star – which is the lowest possible companion mass allowed by the radial velocity measurements - the radius and temperature is approximately equal to 0.7 $R_{\odot}$ and 0.84 $T_{\odot}$, respectively. Using the derived mass and temperature of the primary, we get the flux ratio of the main sequence companion to the primary to be 0.83. So the main sequence companion contributes almost as much as the primary at large wavelengths and the total flux of the system would be twice the expected flux from the primary. We do not see such a behavior in 2MASS or WISE photometry. For higher companion masses, this effect would be more pronounced. Thus, we rule out a main sequence companion for this system.

Depending on the composition, the maximum mass of a white dwarf ranges between 1.35$-$1.40 M$_{\odot}$ \citep{caiazzo_2021}. Fig.~\ref{fig:m1m2} indicates that for nearly edge-on orbits there is a possibility of the companion being a massive WD. For all moderate inclinations we expect the companion to be more massive than any reasonable limit on the WD mass, and therefore it would need to be a neutron star. 

Along with the RVs we also use the lightcurve data to constrain the mass of companion by simulating the lightcurves and RV variation. To simulate the periodic ellipsoidal variation we use $\tt{PHOEBE}$\footnote{\url{http://phoebe-project.org/}} \citep{prsa_modeling_2018}. The stellar and orbital parameters are also taken as inputs to \texttt{PHOEBE} to simulate the lightcurve and the RV variation. To make this process computationally efficient, after phase-folding the lightcurve we bin it into two-hour time bins using $\tt{lightkurve}$ \citep{lightkurve}, reducing the total number of data points from 821 to 36.

If the secondary is a compact object, then for $M_2 \sim$ 0.9 M$_{\odot}$, the upper limit on the radius of the secondary of about 0.008 R$_{\odot}$ can be obtained from the mass-radius relation of WDs \citep{Chandra_2020}, and the flux change due to the eclipse of primary due to secondary would be about 0.02\% for edge-on orbits, which is much smaller than the sensitivity of our data. Therefore, we would not see eclipses even if the system was edge-on, and therefore the lack of eclipses does not provide a constraint on the inclination. The same is true for microlensing as its effect is of the same order as ecplises \citep{Marsh_2001}.
Limb darkening and gravity darkening are second order effects and given the small amplitude of variation and noisiness of our data their effects are nearly negligible as well. Thus we model the surface using the default CK2004 atmospheric models \citep{CK2004} provided in \texttt{PHOEBE}, which includes gravity and limb darkening, even though these effects are small. The two solutions obtained in Sec.~\ref{subsec:Stellar_Param} from fitting the spectroscopic and the photometric data indicated that our target has temperature $T>$ 8000 K. Thus we use a purely radiative atmospheric model with gravity darkening coefficient $\beta$ = 1 \citep{Zeipel_1924, Claret_1999}. We simulate the lightcurve in the ZTF $g$ and $r$ bands and fit them to observed lightcurve. 

Since the effects of limb darkening and gravity darkening are small, the dependence of lightcurve on the temperature is also be minimal, and the \texttt{PHOEBE}-generated lightcurves showed this behavior. Therefore, we fix the temperature of the primary to 8500 K and 10000 K, for two solutions respectively, based on the spectrophotometric value obtained in Sec.~\ref{subsec:Stellar_Param} for ease of computation with little downside. We leave the surface gravity, radius of the primary, mass of the secondary and the inclination as free parameters. In addition to these, we also leave the systemic velocity, $v_{\gamma}$, and initial phase of the orbit, $t_{0,\mathrm{sup-conj}}$, as free parameters.

\subsection{Combined Fit}
\label{subsec:comb_fit}
With all the datasets assembled, we finally perform joint fitting where we add all the log-likelihoods and the associated log-priors, and maximize the resulting likelihood using \texttt{emcee}. Each dataset constrains different stellar and orbital parameters in a distinct way. The spectroscopic fit constrains $T_{\mathrm{eff}}$, $\log g$, [Fe/H] independent of extinction. The lightcurve and RV fits also constrain $\log g$ and radius of the primary independent of extinction, and also constrain $M_{2}$, inclination. The photometric fit is most sensitive to extinction and constrains $T_{\mathrm{eff}}$, $\log g$, [Fe/H], $R$, $E(B-V)$.

We explore two solutions, centered around the two local minima we found from the initial spectrophotometric fitting in Sec.~\ref{subsec:Stellar_Param} separately. The solution with the lower extinction value results in a primary mass of 0.18 M$_{\odot}$ and temperature of $\sim 8500$K, while the solution with higher extinction value results in primary mass of 0.35 M$_{\odot}$ and temperature of $\sim$ 10000 K. Looking at the final fits to the lightcurve and based on the theoretical expectations and empirical data on such binary systems we present the solution with the lower extinction value as the fiducial solution. We explain this choice in Sec.~\ref{subsec:comp}.

\begin{figure*}[ht!]
    \centering
    \includegraphics[scale=0.3]{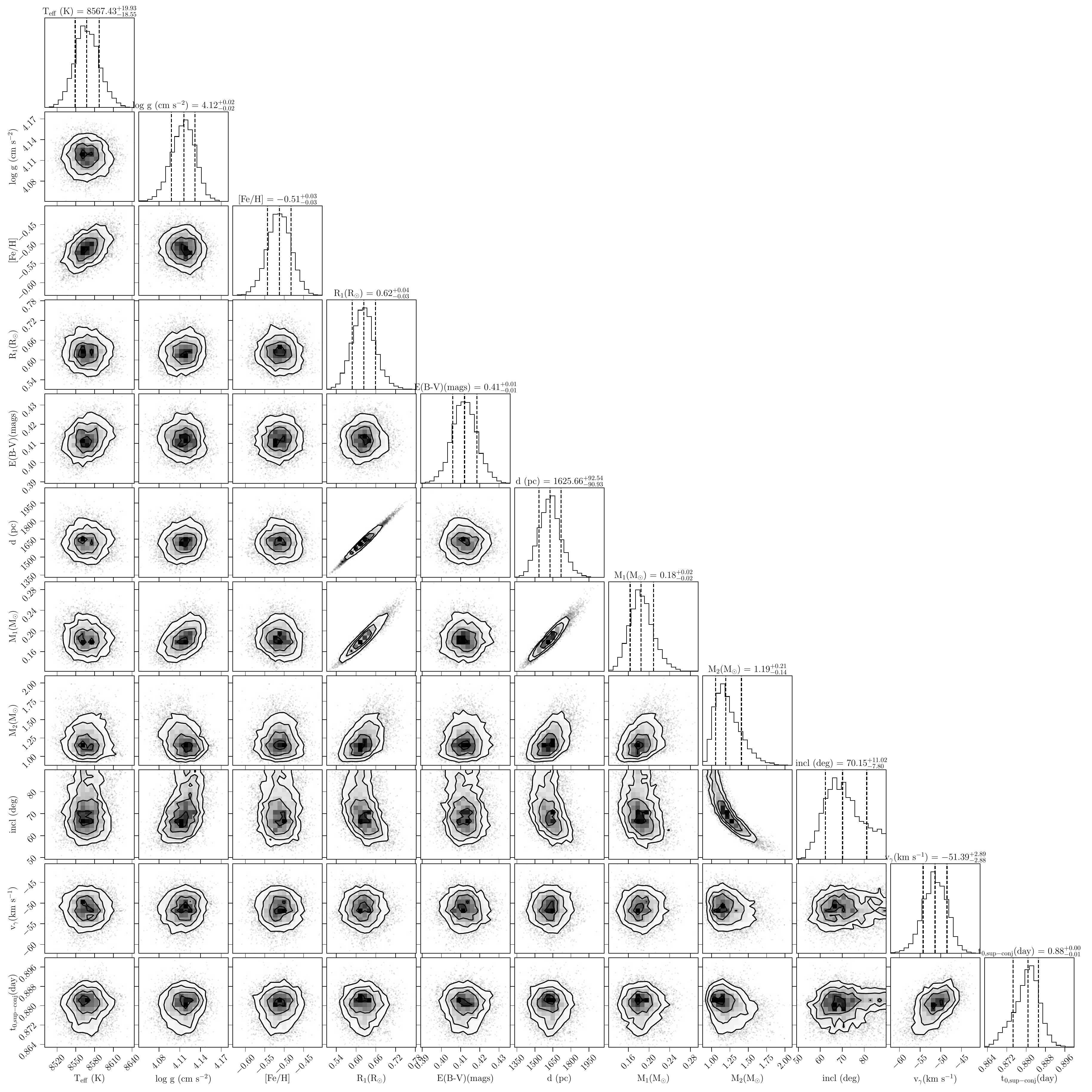}
    \caption{Posterior distribution from the joint parameter fitting of spectroscopic, photometric, RV and lightcurve datasets}
    \label{fig:corner_lowebv}
\end{figure*}

\begin{figure}
    \centering
    \includegraphics[width=\columnwidth]{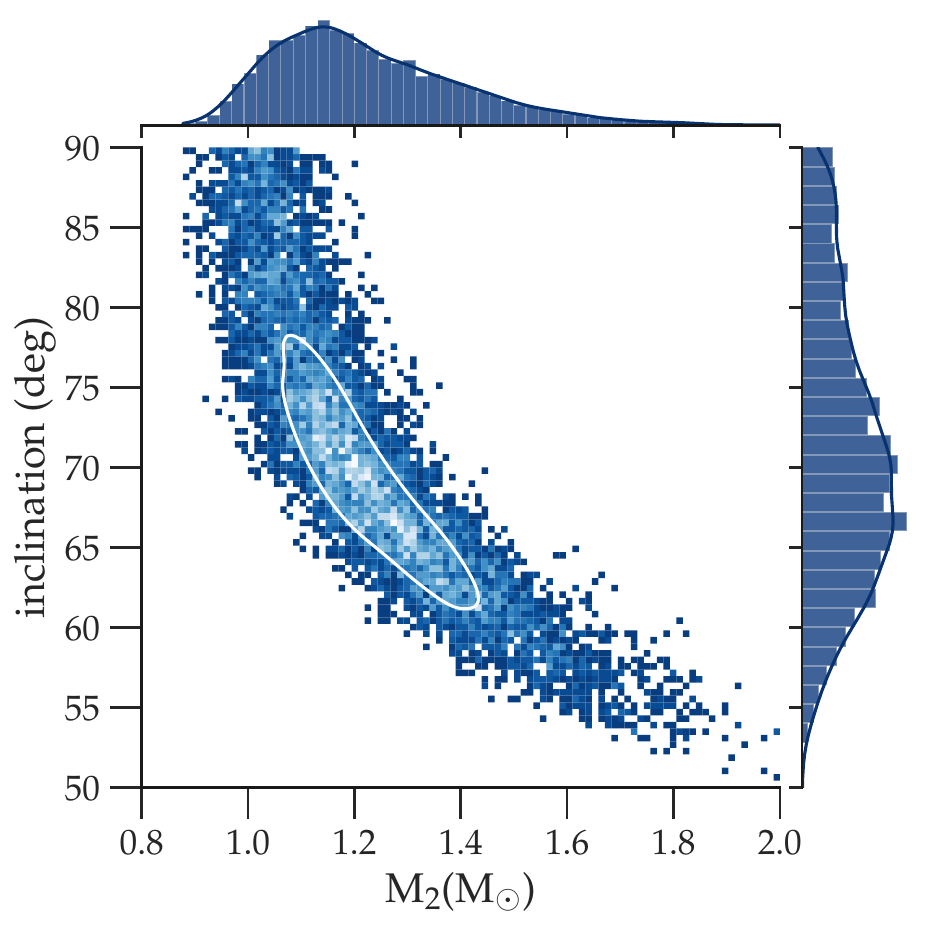}
    \caption{Distribution of $M_2$ against inclination. In white we show the 1-sigma contour.}
    \label{fig:M2}
\end{figure}

The results are tabulated in Table.~\ref{tab:res} and the combined fit is shown in Fig.~\ref{fig:bosz_fit} and Fig.~\ref{fig:rvlc}. The posterior distributions for stellar parameters are shown in Fig. \ref{fig:corner_lowebv}. The mass of the primary is found to be $M_1$ = 0.18 M$_{\odot}$ and the mass of the secondary to be $M_2$ = 1.19 M$_{\odot}$, with 1$\sigma$ upper limit of  1.41 M$_{\odot}$ and 3$\sigma$ lower limit around 0.9 M$_{\odot}$ as shown in Fig.~\ref{fig:M2}, suggesting a massive WD or a neutron star companion. The 1-sigma errors reported are calculated at 16th and 84th percentile of the posterior distributions.

\begin{table}
\begin{center}
    \caption{Observed and derived parameters for SDSS~J0229+7130}
    \begin{threeparttable}
    \begin{tabular}{c c} 
         \hline \hline
         Parameter & Value \\ \hline 
         \textit{Gaia} DR3 Source ID$^*$ & 545437241454018304 \\
         R.A. J2000$^*$ & 02:29:32.29\\
         Dec. J2000$^*$ & +71:30:02.48\\
         G (mag)$^*$ & 16.28\\
         G$_\mathrm{BP}$ - G$_\mathrm{RP} $ (mag) $^*$& 0.77\\
         d (pc) & $1625^{+92}_{-92}$ \\
         \hline \hline
         Orbital Parameters \\ 
         \hline \hline
         Period (hours) & 35.8703 $\pm$ 0.0006 \\
         K (km s$^{-1}$) & 169 $\pm$ 3 \\
         \hline \hline
         Stellar Parameters \\ 
         \hline \hline 
         $\log g _{1}$ & 4.11 $\pm$ 0.01 \\
         $M_1$ (M$_{\odot}$) & 0.18 $\pm$ 0.02 \\
         $M_2$ (M$_{\odot}$) & 1.19$^{+0.21}_{-0.14}$ \\
         $R_1$ (R$_{\odot}$) & 0.62 $\pm$ 0.04  \\
         T$_{\mathrm{eff}}$ (K) & 8567 $\pm$ 20 \textsuperscript{\textdagger}\\
         E(B-V) & 0.41 $\pm$ 0.01 \\
         $[$Fe/H$]$ & -0.51 $\pm$ 0.03\textsuperscript{\textdagger}\\
         i (deg) & $70^{+11}_{-8}$ \\
         \hline \hline
    \end{tabular}
    \begin{tablenotes}
     \item[*] \cite{Gaia_DR3}
     \item[\textdagger] The reported errors are numerical and could be underestimated. A more conservative error estimate would be 100 K and 0.1 dex -- approximately half the BOSZ model grid spacing
   \end{tablenotes}
    \end{threeparttable}
    \label{tab:res}

\end{center}
\end{table}

\section{Discussion}
\label{sec:discussion}

In this paper we present the detection of a pre-ELM WD in a binary with a massive compact companion -- likely a massive white dwarf or a neutron star. We perform a joint analysis of the radial velocity curve and of the photometric lightcurve of the visible pre-ELM WD to determine the allowed range of the masses of companion and find it to be $1.19^{+0.21}_{-0.14}$ M$_\odot$. 

\subsection{Formation scenarios}

The formation of such systems has been theoretically studied in detail as ELM WDs were found in binaries with pulsars and white dwarfs \citep{Liu_2011,Shao_2012,istrate_formation_2014,li_formation_2019}. We follow the various evolutionary tracks derived in these works and piece together the evolutionary pathway of our target. Both ELM WD-WD double degenerate binaries and ELM WD-NS binaries can form via two channels -- stable Roche lobe overflow channel or common envelope ejection channel -- both of which involve the massive compact companion accreting mass from a main sequence star \citep{Nelson_2004,istrate_formation_2014,li_formation_2019}. Following results from \cite{li_formation_2019}, we see that the large period ($>8$ hours) and low mass ELM WDs with WD companions could have formed only via the Roche lobe channel. While the formation channel involving common envelope ejection is not well studied, especially with neutron star companions \citep{Istrate_2016}, common envelope ejection channel is unlikely to produce the orbital period and primary mass combination that we observe. Thus, regardless of whether the unseen companion in our target is a massive white dwarf or a neutron star, we suggest that the system formed via a stable Roche lobe overflow from the main sequence star (which later became the primary pre-ELM WD) to the compact companion. Based on final orbital periods, masses and evolutionary models of \cite{li_formation_2019}, we deduce that the system must have arisen from a binary with initial orbital period of the same order as final orbital period, with the initial mass of primary between 1.2-2 M$_{\odot}$.
\subsection{Nature of the companion}
So far we have guessed the initial properties of the binary regardless of the nature of the companion.
The initial companion mass is sensitive to the nature of the companion. Following \cite{li_formation_2019}, we find that if the companion is a WD, then its initial mass pre-Roche lobe interaction is around 0.6-0.8 M$_{\odot}$. WDs less massive than this range cannot accrete enough mass to grow to the observed mass of $>$ 1M$_{\odot}$ because the mass accumulation is limited by the hydrogen burning rate on the surface of the white dwarf and by the resulting stellar winds. WDs which are more massive than this range cannot appreciably gain mass due to hydrogen-shell flashes causing inefficient mass accumulation. 
If the companion is a neutron star, results from \cite{Shao_2012} suggest that a 1.1-1.2 M$_\odot$ neutron star in a binary system with properties similar to those of our target can accrete about 0.3 M$_{\odot}$ worth of mass over the course of the evolution of ELM WD and lead to a system like ours. While the exact parameter values depend on various factors such as magnetic braking index, metallicity and, more importantly, on the exact nature of the companion, these are the representative values characteristic of systems such as our target and which suggest plausible scenarios such systems could have evolved from. 

If the companion is a neutron star, the accretion of 0.3 M$_{\odot}$ worth of mass would spin up the neutron star and lead to a milli-second pulsar with a spin-period of few milli-seconds \citep{Liu_2011}. The system is not detected in the NRAO VLA Sky Survey (NVSS; \citealt{condon_nrao_1998}) at 1.4 GHz down to a $\sim 2.5$ mJy flux level. X-ray signals can be expected from the surface of pulsars, especially from the magnetosphere of their polar caps in the case of recycled pulsars \citep{Zavlin_2002,Zavlin_2006,kilic_elm_2011}. Looking at data from X-ray survey ROSAT \citep{Truemper_1982} we find no X-ray signal near our target. However, these are all-sky surveys with short exposure times and consequently high flux limits and thus only the brightest pulsars would be visible \citep{Danner_1994,Becker_1996}. Using ROSAT flux limit of $10^{-13}$ erg cm$^{-2}$ s$^{-1}$ and assuming zero extinction in X-ray wavelengths we obtain upper limit on X-ray luminosity to be $L_{\mathrm{X}} < 3.5 \times 10^{31} \mathrm{erg s^{-1}}$. Typical milli-second pulsars have X-ray luminosities below this limit \citep{Bogdanov_2006,Lee_2018} and therefore would remain undetected. The presence of extinction would make their detection in ROSAT even more difficult. 

As has been previously observed, the eccentricity for systems similar to ours is very small \citep{Edwards_2001,istrate_formation_2014}. The binary must have gone through a common envelope phase before the more massive companion formed a compact remnant \citep{li_formation_2019} resulting in a circular orbit. While one would expect that the formation of a neutron star would give a kick to the primary leading to an eccentric orbit, the subsequent Roche-lobe overflow would circularize the orbit, as has been observed for such binaries. The phase-folded RV curve and the lightcurve we obtain are consistent with a circular orbit. In principle, the eccentricity could be left as a free parameter in the analysis. However, given the strong theoretical footing for a nearly zero eccentricity, we do not expect the results to change in a significant way and leave it at zero.

\subsection{Nature of the primary}
While we have assumed that the primary is a pre-ELM WD, we can verify our assumptions using derived parameters. Based on the large observed brightness ($\sim$ 1.5 L$_{\odot}$) and low mass, the primary cannot be a main-sequence star. In addition, the primary is too cold and much less massive than an sdB star ($T_{\mathrm{sdB}}$ $>$ 20000 K and $M_{\mathrm{sdB}}$ $\sim$ 0.5 M$_{\odot}$) \citep{Heber_2009}. We conclude that despite its position on the color-magnitude diagram (Fig. \ref{fig:color-mag}), which is close to the main-sequence track, the primary is indeed a pre-ELM WD. We also see that the primary has an anomalously large luminosity (or, equivalently, a larger radius) compared to pre-ELM sample from \cite{el-badry_birth_2021}, including those which are still mass transferring. This can be explained in the context of the Roche lobe channel as well: our target has a much greater orbital period compared to their sample, and consequently during the binary evolution the Roche-lobe mass transfer is expected to terminate earlier leading to a larger radius. 

\subsection{Comparison with theoretical models}
\label{subsec:comp}
\begin{figure}
    \centering
    \includegraphics[width=\columnwidth]{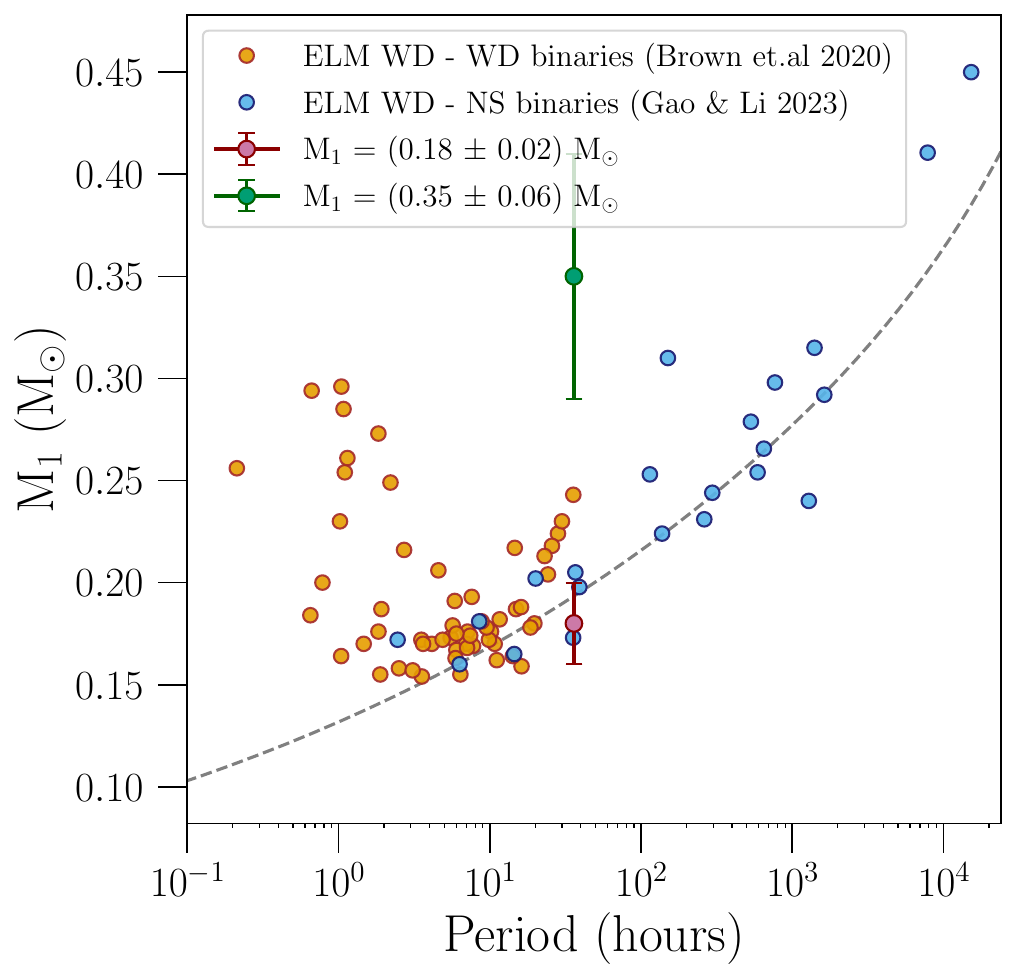}
    \caption{Mass vs. orbital periods for ELM WDs binaries is plotted along with both solutions obtained for our target. Binaries with WD companions are selected from \cite{Brown_2020}, while binaries with NS companions are selected from \cite{Gao_2023}. The dashed line is the theoretical mass-orbital period relation for the Roche lobe channel from \cite{Gao_2023}.}
    \label{fig:M_P}
\end{figure}

The final stage of the close binary evolution through Roche lobe overflow results in a relationship between the primary masses and the orbital periods. These values are shown in Fig.~\ref{fig:M_P}, where we compare our target with ELM WD-WD binaries from the ELM survey \citep{Brown_2020} and and ELM WD-NS binaries from \cite{Gao_2023}. The short-period binaries (P $<$ 8 hrs) with a broad distribution of the primary masses are from the common envelope channel, whereas the Roche lobe channel forms a relatively narrow locus of objects with a strong mass-period dependence. Our target with its period of $P=35.87$ hours falls in the long-period region of the expected Roche lobe channel track. For comparison, we show the theoretically expected mass-period relations from binary simulations consisting of a neutron star and a WD primary. 

\begin{figure}
    \centering
    \includegraphics[width=\columnwidth]{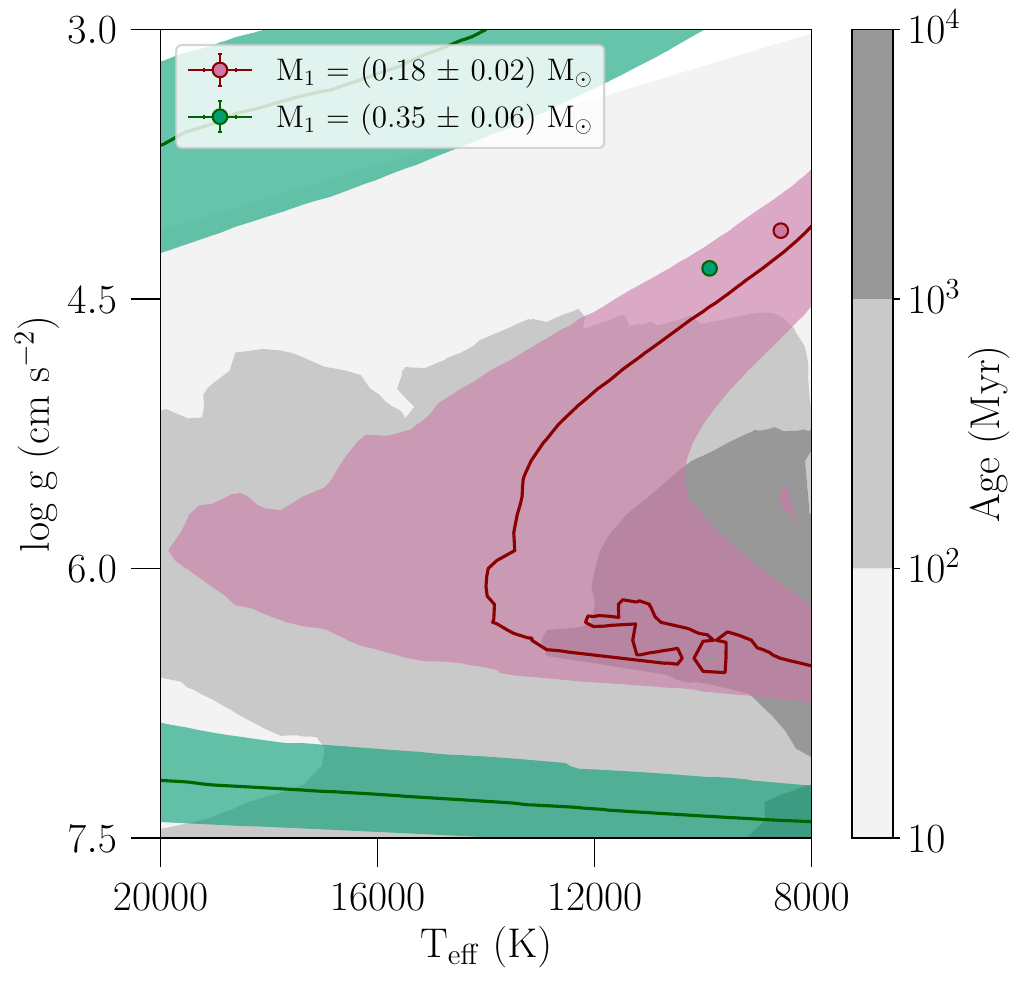}
    \caption{Cooling curves and the ages for ELM WDs from models by \cite{Althaus_2013} is shown in this figure. The pink solid line is the cooling curve for 0.18 M$_{\odot}$ ELM WD and the pink band corresponds to cooling curves for masses in the range of 0.16-0.2 M$_{\odot}$ ELM WDs. The green solid line is the cooling curve for 0.35 M$_{\odot}$ ELM WD and the green band corresponds to cooling curves for masses in the range of 0.29-0.41 M$_{\odot}$ ELM WDs. The results for both the solutions are also plotted. The age ranges for ELM WDs is shown in various shades of gray contours.}
    \label{fig:Cooling}
\end{figure}

In Fig.~\ref{fig:Cooling}, we show the cooling curves and the ages for ELM WDs from \citet{Althaus_2013}. We see that the target is currently younger than 100 Myrs and is in its contraction stage. Using models of \citet{istrate_timescale_2014}, we suggest that the primary will cool down, contract, and settle onto the WD cooling track in a time of the order of 0.2--2 Gyr. The gravitational wave merger timescale for our target is of order a few hundred billion years. Comparing the two time scales we conclude that this system will end up as a stable binary of an ELM WD and a massive compact object whose orbital period will remain at its current value for much longer than the Hubble time.

In Fig.~\ref{fig:M_P} and Fig.~\ref{fig:Cooling}, we plot both obtained solutions. In Fig.~\ref{fig:M_P}, the higher mass solution, shown in green, is above the expected track, while the low mass solution, shown in red, is consistent with the theoretical expectation. In Fig.~\ref{fig:Cooling}, the difference is even more stark - the low mass solution is well within the expected theoretical cooling curves while the high mass solution is clearly inconsistent. Therefore, while both solutions are statistically viable, the solution with the higher primary mass is inconsistent with the astrophysical evolutionary tracks, and we determine the mass of the primary to be $M_1$ = 0.18 M$_{\odot}$. 


\subsection{Sources of uncertainty}


One of the major sources of error in our analysis is the estimation of extinction. As we have discussed, we tried to take into account the various uncertainties associated with it. The uncertainty in extinction along line of sight and the degeneracy between the color excess and other stellar parameters gives two local $\chi^2$ minima in the multi-dimensional parameter space of the fit. The high values of extinction are not ruled out based on the value of the $\chi^2$ or of the likelihood. High extinction solution has higher mass (Sec.~\ref{subsec:comb_fit}), and looking at Fig.~\ref{fig:M_P} and Fig.~\ref{fig:Cooling}, we find this solution (shown in green) is not favored by the models. Following Fig.~\ref{fig:m1m2}, a lower primary mass suggests a lower secondary mass and thus the low extinction solution weakens the case for a neutron star companion. 

In addition to the extinction, the other uncertain part is the polynomial fitting of the continuum. While we have presented results for the polynomial of order 6, we refit the spectrum by increasing the order of the polynomial to make sure this is a stable solution. We find that the final primary mass does not change appreciably and remains within 1-$\sigma$ of our quoted value. We therefore present the results from the fitting with the lowest order polynomial which gives a stable solution. We also performed our analysis by fitting continuum normalized absorption lines alone instead of the full spectral fitting. This technique was sensitive to the choice of absorption lines and to the the continuum normalization. We did manage to recover our solution -- albeit with a higher uncertainty, making it less reliable than our fiducial solution.

\section*{Acknowledgments}

G.A.P. thanks W. Brown for useful comments. 
V.C. gratefully acknowledges a Peirce Fellowship from Harvard University. 
N.L.Z. acknowledges support by the JHU President's Frontier Award and by the seed grant from the JHU Institute for Data Intensive Engineering and Science.
B.G. received funding from the European Research Council (ERC) under the European Union’s Horizon 2020 research and innovation programme (Grant agreement No. 101020057).
Y.Z. gratefully acknowledge support from NASA grants NNH17ZDA001N, 80NSSC22K0494, 80NSSC21K0242 and 80NSSC19K0112.
N.R.C. is supported by the National Science Foundation Graduate Research Fellowship Program under Grant No. DGE2139757. Any opinions, findings, and conclusions or recommendations expressed in this material are those of the authors and do not necessarily reflect the views of the National Science Foundation.

Based on observations obtained with the Samuel Oschin 48-inch Telescope at the Palomar Observatory as part of the Zwicky Transient Facility project. ZTF is supported by the National Science Foundation under Grant No. AST-1440341 and a
collaboration including Caltech, IPAC, the Weizmann Institute for Science, the Oskar Klein Center at Stockholm University, the University of Maryland, the University of Washington, Deutsches Elektronen-Synchrotron and Humboldt University, Los Alamos National Laboratories, the TANGO Consortium of Taiwan, the University of Wisconsin at Milwaukee, and Lawrence Berkeley National Laboratories. Operations are conducted by COO, IPAC, and UW.

This publication makes use of data products from the Two Micron All Sky Survey, which is a joint project of the University of Massachusetts and the Infrared Processing and Analysis Center/California Institute of Technology, funded by the National Aeronautics and Space Administration and the National Science Foundation. 

Funding for the Sloan Digital Sky Survey V has been provided by the Alfred P. Sloan Foundation, the Heising-Simons Foundation, the National Science Foundation, and the Participating Institutions. SDSS acknowledges support and resources from the Center for High-Performance Computing at the University of Utah. SDSS telescopes are located at Apache Point Observatory, funded by the Astrophysical Research Consortium and operated by New Mexico State University, and at Las Campanas Observatory, operated by the Carnegie Institution for Science. The SDSS web site is \url{www.sdss.org}.

SDSS is managed by the Astrophysical Research Consortium for the Participating Institutions of the SDSS Collaboration, including Caltech, The Carnegie Institution for Science, Chilean National Time Allocation Committee (CNTAC) ratified researchers, The Flatiron Institute, the Gotham Participation Group, Harvard University, Heidelberg University, The Johns Hopkins University, L’Ecole polytechnique f\'{e}d\'{e}rale de Lausanne (EPFL), Leibniz-Institut f\"{u}r Astrophysik Potsdam (AIP), Max-Planck-Institut f\"{u}r Astronomie (MPIA Heidelberg), Max-Planck-Institut f\"{u}r Extraterrestrische Physik (MPE), Nanjing University, National Astronomical Observatories of China (NAOC), New Mexico State University, The Ohio State University, Pennsylvania State University, Smithsonian Astrophysical Observatory, Space Telescope Science Institute (STScI), the Stellar Astrophysics Participation Group, Universidad Nacional Aut\'{o}noma de M\'{e}xico, University of Arizona, University of Colorado Boulder, University of Illinois at Urbana-Champaign, University of Toronto, University of Utah, University of Virginia, Yale University, and Yunnan University.


\software{astropy \citep{astropy:2013, astropy:2018, astropy:2022}, numpy \citep{numpy}, scipy \citep{scipy}, matplotlib \citep{matplotlib}, lightkurve \citep{lightkurve}, eleanor \citep{eleanor_2019}, PHOEBE \citep{prsa_modeling_2018}}

\section*{Data availability}
Photometric data from GALEX, SDSS, Pan-STARRS, 2MASS, Cat-WISE used in this paper are publicly available from the corresponding archives. The ZTF and TESS lightcurve data used is also publicly available \citep{ZTF_doi,TESS_18_DOI,TESS_19_DOI}. The SDSS-V data are not public but the reduced and calibrated data are made available as electronic supplement to the paper. The APO data used in this paper are also made available as electronic supplement to the paper.








\bibliography{Main}{}

\begin{thebibliography}{}
\expandafter\ifx\csname natexlab\endcsname\relax\def\natexlab#1{#1}\fi
\providecommand{\url}[1]{\href{#1}{#1}}
\providecommand{\dodoi}[1]{doi:~\href{http://doi.org/#1}{\nolinkurl{#1}}}
\providecommand{\doeprint}[1]{\href{http://ascl.net/#1}{\nolinkurl{http://ascl.net/#1}}}
\providecommand{\doarXiv}[1]{\href{https://arxiv.org/abs/#1}{\nolinkurl{https://arxiv.org/abs/#1}}}

\bibitem[{{Ahumada} {et~al.}(2020){Ahumada}, {Allende Prieto}, {Almeida},
  {Anders}, {Anderson}, {Andrews}, {Anguiano}, {Arcodia}, {Armengaud},
  {Aubert}, {Avila}, {Avila-Reese}, {Badenes}, {Balland}, {Barger},
  {Barrera-Ballesteros}, {Basu}, {Bautista}, {Beaton}, {Beers}, {Benavides},
  {Bender}, {Bernardi}, {Bershady}, {Beutler}, {Bidin}, {Bird}, {Bizyaev},
  {Blanc}, {Blanton}, {Boquien}, {Borissova}, {Bovy}, {Brandt}, {Brinkmann},
  {Brownstein}, {Bundy}, {Bureau}, {Burgasser}, {Burtin}, {Cano-D{\'\i}az},
  {Capasso}, {Cappellari}, {Carrera}, {Chabanier}, {Chaplin}, {Chapman},
  {Cherinka}, {Chiappini}, {Doohyun Choi}, {Chojnowski}, {Chung}, {Clerc},
  {Coffey}, {Comerford}, {Comparat}, {da Costa}, {Cousinou}, {Covey}, {Crane},
  {Cunha}, {Ilha}, {Dai}, {Damsted}, {Darling}, {Davidson}, {Davies}, {Dawson},
  {De}, {de la Macorra}, {De Lee}, {Queiroz}, {Deconto Machado}, {de la Torre},
  {Dell'Agli}, {du Mas des Bourboux}, {Diamond-Stanic}, {Dillon}, {Donor},
  {Drory}, {Duckworth}, {Dwelly}, {Ebelke}, {Eftekharzadeh}, {Davis Eigenbrot},
  {Elsworth}, {Eracleous}, {Erfanianfar}, {Escoffier}, {Fan}, {Farr},
  {Fern{\'a}ndez-Trincado}, {Feuillet}, {Finoguenov}, {Fofie},
  {Fraser-McKelvie}, {Frinchaboy}, {Fromenteau}, {Fu}, {Galbany}, {Garcia},
  {Garc{\'\i}a-Hern{\'a}ndez}, {Garma Oehmichen}, {Ge}, {Geimba Maia},
  {Geisler}, {Gelfand}, {Goddy}, {Gonzalez-Perez}, {Grabowski}, {Green},
  {Grier}, {Guo}, {Guy}, {Harding}, {Hasselquist}, {Hawken}, {Hayes}, {Hearty},
  {Hekker}, {Hogg}, {Holtzman}, {Horta}, {Hou}, {Hsieh}, {Huber}, {Hunt}, {Ider
  Chitham}, {Imig}, {Jaber}, {Jimenez Angel}, {Johnson}, {Jones},
  {J{\"o}nsson}, {Jullo}, {Kim}, {Kinemuchi}, {Kirkpatrick}, {Kite}, {Klaene},
  {Kneib}, {Kollmeier}, {Kong}, {Kounkel}, {Krishnarao}, {Lacerna}, {Lan},
  {Lane}, {Law}, {Le Goff}, {Leung}, {Lewis}, {Li}, {Lian}, {Lin}, {Long},
  {Longa-Pe{\~n}a}, {Lundgren}, {Lyke}, {Mackereth}, {MacLeod}, {Majewski},
  {Manchado}, {Maraston}, {Martini}, {Masseron}, {Masters}, {Mathur},
  {McDermid}, {Merloni}, {Merrifield}, {M{\'e}sz{\'a}ros}, {Miglio}, {Minniti},
  {Minsley}, {Miyaji}, {Mohammad}, {Mosser}, {Mueller}, {Muna},
  {Mu{\~n}oz-Guti{\'e}rrez}, {Myers}, {Nadathur}, {Nair}, {Nandra}, {Correa do
  Nascimento}, {Nevin}, {Newman}, {Nidever}, {Nitschelm}, {Noterdaeme},
  {O'Connell}, {Olmstead}, {Oravetz}, {Oravetz}, {Osorio}, {Pace}, {Padilla},
  {Palanque-Delabrouille}, {Palicio}, {Pan}, {Pan}, {Parker}, {Paviot},
  {Peirani}, {Ram{\'r}ez}, {Penny}, {Percival}, {Perez-Fournon},
  {P{\'e}rez-R{\`a}fols}, {Petitjean}, {Pieri}, {Pinsonneault}, {Poovelil},
  {Povick}, {Prakash}, {Price-Whelan}, {Raddick}, {Raichoor}, {Ray}, {Rembold},
  {Rezaie}, {Riffel}, {Riffel}, {Rix}, {Robin}, {Roman-Lopes},
  {Rom{\'a}n-Z{\'u}{\~n}iga}, {Rose}, {Ross}, {Rossi}, {Rowlands}, {Rubin},
  {Salvato}, {S{\'a}nchez}, {S{\'a}nchez-Menguiano}, {S{\'a}nchez-Gallego},
  {Sayres}, {Schaefer}, {Schiavon}, {Schimoia}, {Schlafly}, {Schlegel},
  {Schneider}, {Schultheis}, {Schwope}, {Seo}, {Serenelli}, {Shafieloo},
  {Shamsi}, {Shao}, {Shen}, {Shetrone}, {Shirley}, {Silva Aguirre}, {Simon},
  {Skrutskie}, {Slosar}, {Smethurst}, {Sobeck}, {Sodi}, {Souto}, {Stark},
  {Stassun}, {Steinmetz}, {Stello}, {Stermer}, {Storchi-Bergmann},
  {Streblyanska}, {Stringfellow}, {Stutz}, {Su{\'a}rez}, {Sun},
  {Taghizadeh-Popp}, {Talbot}, {Tayar}, {Thakar}, {Theriault}, {Thomas},
  {Thomas}, {Tinker}, {Tojeiro}, {Toledo}, {Tremonti}, {Troup}, {Tuttle},
  {Unda-Sanzana}, {Valentini}, {Vargas-Gonz{\'a}lez}, {Vargas-Maga{\~n}a},
  {V{\'a}zquez-Mata}, {Vivek}, {Wake}, {Wang}, {Weaver}, {Weijmans}, {Wild},
  {Wilson}, {Wilson}, {Wolthuis}, {Wood-Vasey}, {Yan}, {Yang}, {Y{\`e}che},
  {Zamora}, {Zarrouk}, {Zasowski}, {Zhang}, {Zhao}, {Zhao}, {Zheng}, {Zheng},
  {Zhu}, \& {Zou}}]{SDSS_DR16}
{Ahumada}, R., {Allende Prieto}, C., {Almeida}, A., {et~al.} 2020, \apjs, 249,
  3, \dodoi{10.3847/1538-4365/ab929e}

\bibitem[{{Althaus} {et~al.}(2013){Althaus}, {Miller Bertolami}, \&
  {C{\'o}rsico}}]{Althaus_2013}
{Althaus}, L.~G., {Miller Bertolami}, M.~M., \& {C{\'o}rsico}, A.~H. 2013,
  \aap, 557, A19, \dodoi{10.1051/0004-6361/201321868}

\bibitem[{{Antoniadis} {et~al.}(2013){Antoniadis}, {Freire}, {Wex}, {Tauris},
  {Lynch}, {van Kerkwijk}, {Kramer}, {Bassa}, {Dhillon}, {Driebe}, {Hessels},
  {Kaspi}, {Kondratiev}, {Langer}, {Marsh}, {McLaughlin}, {Pennucci}, {Ransom},
  {Stairs}, {van Leeuwen}, {Verbiest}, \& {Whelan}}]{Antoniadis_2013}
{Antoniadis}, J., {Freire}, P. C.~C., {Wex}, N., {et~al.} 2013, Science, 340,
  448, \dodoi{10.1126/science.1233232}

\bibitem[{{Astropy Collaboration} {et~al.}(2013){Astropy Collaboration},
  {Robitaille}, {Tollerud}, {Greenfield}, {Droettboom}, {Bray}, {Aldcroft},
  {Davis}, {Ginsburg}, {Price-Whelan}, {Kerzendorf}, {Conley}, {Crighton},
  {Barbary}, {Muna}, {Ferguson}, {Grollier}, {Parikh}, {Nair}, {Unther},
  {Deil}, {Woillez}, {Conseil}, {Kramer}, {Turner}, {Singer}, {Fox}, {Weaver},
  {Zabalza}, {Edwards}, {Azalee Bostroem}, {Burke}, {Casey}, {Crawford},
  {Dencheva}, {Ely}, {Jenness}, {Labrie}, {Lim}, {Pierfederici}, {Pontzen},
  {Ptak}, {Refsdal}, {Servillat}, \& {Streicher}}]{astropy:2013}
{Astropy Collaboration}, {Robitaille}, T.~P., {Tollerud}, E.~J., {et~al.} 2013,
  \aap, 558, A33, \dodoi{10.1051/0004-6361/201322068}

\bibitem[{{Astropy Collaboration} {et~al.}(2018){Astropy Collaboration},
  {Price-Whelan}, {Sip{\H{o}}cz}, {G{\"u}nther}, {Lim}, {Crawford}, {Conseil},
  {Shupe}, {Craig}, {Dencheva}, {Ginsburg}, {Vand erPlas}, {Bradley},
  {P{\'e}rez-Su{\'a}rez}, {de Val-Borro}, {Aldcroft}, {Cruz}, {Robitaille},
  {Tollerud}, {Ardelean}, {Babej}, {Bach}, {Bachetti}, {Bakanov}, {Bamford},
  {Barentsen}, {Barmby}, {Baumbach}, {Berry}, {Biscani}, {Boquien}, {Bostroem},
  {Bouma}, {Brammer}, {Bray}, {Breytenbach}, {Buddelmeijer}, {Burke},
  {Calderone}, {Cano Rodr{\'\i}guez}, {Cara}, {Cardoso}, {Cheedella}, {Copin},
  {Corrales}, {Crichton}, {D'Avella}, {Deil}, {Depagne}, {Dietrich}, {Donath},
  {Droettboom}, {Earl}, {Erben}, {Fabbro}, {Ferreira}, {Finethy}, {Fox},
  {Garrison}, {Gibbons}, {Goldstein}, {Gommers}, {Greco}, {Greenfield},
  {Groener}, {Grollier}, {Hagen}, {Hirst}, {Homeier}, {Horton}, {Hosseinzadeh},
  {Hu}, {Hunkeler}, {Ivezi{\'c}}, {Jain}, {Jenness}, {Kanarek}, {Kendrew},
  {Kern}, {Kerzendorf}, {Khvalko}, {King}, {Kirkby}, {Kulkarni}, {Kumar},
  {Lee}, {Lenz}, {Littlefair}, {Ma}, {Macleod}, {Mastropietro}, {McCully},
  {Montagnac}, {Morris}, {Mueller}, {Mumford}, {Muna}, {Murphy}, {Nelson},
  {Nguyen}, {Ninan}, {N{\"o}the}, {Ogaz}, {Oh}, {Parejko}, {Parley}, {Pascual},
  {Patil}, {Patil}, {Plunkett}, {Prochaska}, {Rastogi}, {Reddy Janga},
  {Sabater}, {Sakurikar}, {Seifert}, {Sherbert}, {Sherwood-Taylor}, {Shih},
  {Sick}, {Silbiger}, {Singanamalla}, {Singer}, {Sladen}, {Sooley},
  {Sornarajah}, {Streicher}, {Teuben}, {Thomas}, {Tremblay}, {Turner},
  {Terr{\'o}n}, {van Kerkwijk}, {de la Vega}, {Watkins}, {Weaver}, {Whitmore},
  {Woillez}, {Zabalza}, \& {Astropy Contributors}}]{astropy:2018}
{Astropy Collaboration}, {Price-Whelan}, A.~M., {Sip{\H{o}}cz}, B.~M., {et~al.}
  2018, \aj, 156, 123, \dodoi{10.3847/1538-3881/aabc4f}

\bibitem[{{Astropy Collaboration} {et~al.}(2022){Astropy Collaboration},
  {Price-Whelan}, {Lim}, {Earl}, {Starkman}, {Bradley}, {Shupe}, {Patil},
  {Corrales}, {Brasseur}, {N{"o}the}, {Donath}, {Tollerud}, {Morris},
  {Ginsburg}, {Vaher}, {Weaver}, {Tocknell}, {Jamieson}, {van Kerkwijk},
  {Robitaille}, {Merry}, {Bachetti}, {G{"u}nther}, {Aldcroft},
  {Alvarado-Montes}, {Archibald}, {B{'o}di}, {Bapat}, {Barentsen}, {Baz{'a}n},
  {Biswas}, {Boquien}, {Burke}, {Cara}, {Cara}, {Conroy}, {Conseil}, {Craig},
  {Cross}, {Cruz}, {D'Eugenio}, {Dencheva}, {Devillepoix}, {Dietrich},
  {Eigenbrot}, {Erben}, {Ferreira}, {Foreman-Mackey}, {Fox}, {Freij}, {Garg},
  {Geda}, {Glattly}, {Gondhalekar}, {Gordon}, {Grant}, {Greenfield}, {Groener},
  {Guest}, {Gurovich}, {Handberg}, {Hart}, {Hatfield-Dodds}, {Homeier},
  {Hosseinzadeh}, {Jenness}, {Jones}, {Joseph}, {Kalmbach}, {Karamehmetoglu},
  {Ka{l}uszy{'n}ski}, {Kelley}, {Kern}, {Kerzendorf}, {Koch}, {Kulumani},
  {Lee}, {Ly}, {Ma}, {MacBride}, {Maljaars}, {Muna}, {Murphy}, {Norman},
  {O'Steen}, {Oman}, {Pacifici}, {Pascual}, {Pascual-Granado}, {Patil},
  {Perren}, {Pickering}, {Rastogi}, {Roulston}, {Ryan}, {Rykoff}, {Sabater},
  {Sakurikar}, {Salgado}, {Sanghi}, {Saunders}, {Savchenko}, {Schwardt},
  {Seifert-Eckert}, {Shih}, {Jain}, {Shukla}, {Sick}, {Simpson},
  {Singanamalla}, {Singer}, {Singhal}, {Sinha}, {Sip{H{o}}cz}, {Spitler},
  {Stansby}, {Streicher}, {{{S}}umak}, {Swinbank}, {Taranu}, {Tewary},
  {Tremblay}, {Val-Borro}, {Van Kooten}, {Vasovi{'c}}, {Verma}, {de Miranda
  Cardoso}, {Williams}, {Wilson}, {Winkel}, {Wood-Vasey}, {Xue}, {Yoachim},
  {Zhang}, {Zonca}, \& {Astropy Project Contributors}}]{astropy:2022}
{Astropy Collaboration}, {Price-Whelan}, A.~M., {Lim}, P.~L., {et~al.} 2022,
  apj, 935, 167, \dodoi{10.3847/1538-4357/ac7c74}

\bibitem[{{Bailer-Jones} {et~al.}(2021){Bailer-Jones}, {Rybizki}, {Fouesneau},
  {Demleitner}, \& {Andrae}}]{Bailer-Jones_2021}
{Bailer-Jones}, C.~A.~L., {Rybizki}, J., {Fouesneau}, M., {Demleitner}, M., \&
  {Andrae}, R. 2021, \aj, 161, 147, \dodoi{10.3847/1538-3881/abd806}

\bibitem[{{Becker} {et~al.}(1996){Becker}, {Trumper}, {Lundgren}, {Cordes}, \&
  {Zepka}}]{Becker_1996}
{Becker}, W., {Trumper}, J., {Lundgren}, S.~C., {Cordes}, J.~M., \& {Zepka},
  A.~F. 1996, \mnras, 282, L33, \dodoi{10.1093/mnras/282.3.L33}

\bibitem[{{Bianchi} {et~al.}(2017){Bianchi}, {Shiao}, \&
  {Thilker}}]{GALEX_2017}
{Bianchi}, L., {Shiao}, B., \& {Thilker}, D. 2017, \apjs, 230, 24,
  \dodoi{10.3847/1538-4365/aa7053}

\bibitem[{{Blanton} \& {Roweis}(2007)}]{Blanton_2007}
{Blanton}, M.~R., \& {Roweis}, S. 2007, \aj, 133, 734, \dodoi{10.1086/510127}

\bibitem[{{Bobrick} {et~al.}(2017){Bobrick}, {Davies}, \&
  {Church}}]{Bobrick_2017}
{Bobrick}, A., {Davies}, M.~B., \& {Church}, R.~P. 2017, \mnras, 467, 3556,
  \dodoi{10.1093/mnras/stx312}

\bibitem[{{Bobrick} {et~al.}(2022){Bobrick}, {Zenati}, {Perets}, {Davies}, \&
  {Church}}]{Bobrick_2022}
{Bobrick}, A., {Zenati}, Y., {Perets}, H.~B., {Davies}, M.~B., \& {Church}, R.
  2022, \mnras, 510, 3758, \dodoi{10.1093/mnras/stab3574}

\bibitem[{{Bogdanov} {et~al.}(2006){Bogdanov}, {Grindlay}, {Heinke}, {Camilo},
  {Freire}, \& {Becker}}]{Bogdanov_2006}
{Bogdanov}, S., {Grindlay}, J.~E., {Heinke}, C.~O., {et~al.} 2006, \apj, 646,
  1104, \dodoi{10.1086/505133}

\bibitem[{Bohlin {et~al.}(2017)Bohlin, Mészáros, Fleming, Gordon, Koekemoer,
  \& Kovács}]{bohlin_new_2017}
Bohlin, R.~C., Mészáros, S., Fleming, S.~W., {et~al.} 2017, AJ, 153, 234,
  \dodoi{10.3847/1538-3881/aa6ba9}

\bibitem[{{Bowen} \& {Vaughan}(1973)}]{bowen_1973}
{Bowen}, I.~S., \& {Vaughan}, A.~H., J. 1973, \ao, 12, 1430,
  \dodoi{10.1364/AO.12.001430}

\bibitem[{{Brasseur} {et~al.}(2019){Brasseur}, {Phillip}, {Fleming},
  {Mullally}, \& {White}}]{astrocut_2019}
{Brasseur}, C.~E., {Phillip}, C., {Fleming}, S.~W., {Mullally}, S.~E., \&
  {White}, R.~L. 2019, {Astrocut: Tools for creating cutouts of TESS images},
  Astrophysics Source Code Library, record ascl:1905.007.
\newblock \doeprint{1905.007}

\bibitem[{{Brown} {et~al.}(2010){Brown}, {Kilic}, {Allende Prieto}, \&
  {Kenyon}}]{Brown_2010}
{Brown}, W.~R., {Kilic}, M., {Allende Prieto}, C., \& {Kenyon}, S.~J. 2010,
  \apj, 723, 1072, \dodoi{10.1088/0004-637X/723/2/1072}

\bibitem[{Brown {et~al.}(2020)Brown, Kilic, Kosakowski, Andrews, Heinke,
  Agüeros, Camilo, Gianninas, Hermes, \& Kenyon}]{Brown_2020}
Brown, W.~R., Kilic, M., Kosakowski, A., {et~al.} 2020, The Astrophysical
  Journal, 889, 49, \dodoi{10.3847/1538-4357/ab63cd}

\bibitem[{{Caiazzo} {et~al.}(2021){Caiazzo}, {Burdge}, {Fuller}, {Heyl},
  {Kulkarni}, {Prince}, {Richer}, {Schwab}, {Andreoni}, {Bellm}, {Drake},
  {Duev}, {Graham}, {Helou}, {Mahabal}, {Masci}, {Smith}, \&
  {Soumagnac}}]{caiazzo_2021}
{Caiazzo}, I., {Burdge}, K.~B., {Fuller}, J., {et~al.} 2021, \nat, 595, 39,
  \dodoi{10.1038/s41586-021-03615-y}

\bibitem[{{Capitanio} {et~al.}(2017){Capitanio}, {Lallement}, {Vergely},
  {Elyajouri}, \& {Monreal-Ibero}}]{Capitanio_2017}
{Capitanio}, L., {Lallement}, R., {Vergely}, J.~L., {Elyajouri}, M., \&
  {Monreal-Ibero}, A. 2017, \aap, 606, A65, \dodoi{10.1051/0004-6361/201730831}

\bibitem[{{Castelli} \& {Kurucz}(2003)}]{CK2004}
{Castelli}, F., \& {Kurucz}, R.~L. 2003, in Modelling of Stellar Atmospheres,
  ed. N.~{Piskunov}, W.~W. {Weiss}, \& D.~F. {Gray}, Vol. 210, A20,
  \dodoi{10.48550/arXiv.astro-ph/0405087}

\bibitem[{{Chandra} {et~al.}(2020){Chandra}, {Hwang}, {Zakamska}, \&
  {Cheng}}]{Chandra_2020}
{Chandra}, V., {Hwang}, H.-C., {Zakamska}, N.~L., \& {Cheng}, S. 2020, \apj,
  899, 146, \dodoi{10.3847/1538-4357/aba8a2}

\bibitem[{{Claret}(1999)}]{Claret_1999}
{Claret}, A. 1999, in Astronomical Society of the Pacific Conference Series,
  Vol. 173, Stellar Structure: Theory and Test of Connective Energy Transport,
  ed. A.~{Gimenez}, E.~F. {Guinan}, \& B.~{Montesinos}, 277

\bibitem[{Condon {et~al.}(1998)Condon, Cotton, Greisen, Yin, Perley, Taylor, \&
  Broderick}]{condon_nrao_1998}
Condon, J.~J., Cotton, W.~D., Greisen, E.~W., {et~al.} 1998, The Astronomical
  Journal, 115, 1693, \dodoi{10.1086/300337}

\bibitem[{Corrales(2015)}]{lia_corrales_2015_15991}
Corrales, L. 2015, {dust: Calculate the intensity of dust scattering halos in
  the X-ray}, 1.0,  Zenodo, \dodoi{10.5281/zenodo.15991}

\bibitem[{{Danner} {et~al.}(1994){Danner}, {Kulkarni}, \&
  {Thorsett}}]{Danner_1994}
{Danner}, R., {Kulkarni}, S.~R., \& {Thorsett}, S.~E. 1994, \apjl, 436, L153,
  \dodoi{10.1086/187655}

\bibitem[{{Edwards} \& {Bailes}(2001)}]{Edwards_2001}
{Edwards}, R.~T., \& {Bailes}, M. 2001, \apjl, 547, L37, \dodoi{10.1086/318893}

\bibitem[{{Eggleton}(1983)}]{Eggleton_1983}
{Eggleton}, P.~P. 1983, \apj, 268, 368, \dodoi{10.1086/160960}

\bibitem[{{Eisenhardt} {et~al.}(2020){Eisenhardt}, {Marocco}, {Fowler},
  {Meisner}, {Kirkpatrick}, {Garcia}, {Jarrett}, {Koontz}, {Marchese},
  {Stanford}, {Caselden}, {Cushing}, {Cutri}, {Faherty}, {Gelino}, {Gonzalez},
  {Mainzer}, {Mobasher}, {Schlegel}, {Stern}, {Teplitz}, \&
  {Wright}}]{CatWISE_2020}
{Eisenhardt}, P. R.~M., {Marocco}, F., {Fowler}, J.~W., {et~al.} 2020, \apjs,
  247, 69, \dodoi{10.3847/1538-4365/ab7f2a}

\bibitem[{El-Badry {et~al.}(2021)El-Badry, Rix, Quataert, Kupfer, \&
  Shen}]{el-badry_birth_2021}
El-Badry, K., Rix, H.-W., Quataert, E., Kupfer, T., \& Shen, K.~J. 2021,
  Monthly Notices of the Royal Astronomical Society, 508, 4106,
  \dodoi{10.1093/mnras/stab2583}

\bibitem[{{Feinstein} {et~al.}(2019){Feinstein}, {Montet}, {Foreman-Mackey},
  {Bedell}, {Saunders}, {Bean}, {Christiansen}, {Hedges}, {Luger}, {Scolnic},
  \& {Cardoso}}]{eleanor_2019}
{Feinstein}, A.~D., {Montet}, B.~T., {Foreman-Mackey}, D., {et~al.} 2019,
  \pasp, 131, 094502, \dodoi{10.1088/1538-3873/ab291c}

\bibitem[{{Fitzpatrick}(1999)}]{Fitzpatrick_1999}
{Fitzpatrick}, E.~L. 1999, \pasp, 111, 63, \dodoi{10.1086/316293}

\bibitem[{{Flewelling}(2018)}]{PanSTARRS_2018}
{Flewelling}, H. 2018, in American Astronomical Society Meeting Abstracts, Vol.
  231, American Astronomical Society Meeting Abstracts \#231, 436.01

\bibitem[{Foreman-Mackey {et~al.}(2013)Foreman-Mackey, Hogg, Lang, \&
  Goodman}]{Foreman_Mackey_2013}
Foreman-Mackey, D., Hogg, D.~W., Lang, D., \& Goodman, J. 2013, Publications of
  the Astronomical Society of the Pacific, 125, 306, \dodoi{10.1086/670067}

\bibitem[{Fryer {et~al.}(1999)Fryer, Woosley, Herant, \&
  Davies}]{fryer_merging_1999}
Fryer, C.~L., Woosley, S.~E., Herant, M., \& Davies, M.~B. 1999, ApJ, 520, 650,
  \dodoi{10.1086/307467}

\bibitem[{{Gaia Collaboration} {et~al.}(2023){Gaia Collaboration}, {Vallenari},
  {Brown}, {Prusti}, {de Bruijne}, {Arenou}, {Babusiaux}, {Biermann},
  {Creevey}, {Ducourant}, {Evans}, {Eyer}, {Guerra}, {Hutton}, {Jordi},
  {Klioner}, {Lammers}, {Lindegren}, {Luri}, {Mignard}, {Panem}, {Pourbaix},
  {Randich}, {Sartoretti}, {Soubiran}, {Tanga}, {Walton}, {Bailer-Jones},
  {Bastian}, {Drimmel}, {Jansen}, {Katz}, {Lattanzi}, {van Leeuwen}, {Bakker},
  {Cacciari}, {Casta{\~n}eda}, {De Angeli}, {Fabricius}, {Fouesneau},
  {Fr{\'e}mat}, {Galluccio}, {Guerrier}, {Heiter}, {Masana}, {Messineo},
  {Mowlavi}, {Nicolas}, {Nienartowicz}, {Pailler}, {Panuzzo}, {Riclet}, {Roux},
  {Seabroke}, {Sordo}, {Th{\'e}venin}, {Gracia-Abril}, {Portell}, {Teyssier},
  {Altmann}, {Andrae}, {Audard}, {Bellas-Velidis}, {Benson}, {Berthier},
  {Blomme}, {Burgess}, {Busonero}, {Busso}, {C{\'a}novas}, {Carry}, {Cellino},
  {Cheek}, {Clementini}, {Damerdji}, {Davidson}, {de Teodoro}, {Nu{\~n}ez
  Campos}, {Delchambre}, {Dell'Oro}, {Esquej}, {Fern{\'a}ndez-Hern{\'a}ndez},
  {Fraile}, {Garabato}, {Garc{\'\i}a-Lario}, {Gosset}, {Haigron}, {Halbwachs},
  {Hambly}, {Harrison}, {Hern{\'a}ndez}, {Hestroffer}, {Hodgkin}, {Holl},
  {Jan{\ss}en}, {Jevardat de Fombelle}, {Jordan}, {Krone-Martins}, {Lanzafame},
  {L{\"o}ffler}, {Marchal}, {Marrese}, {Moitinho}, {Muinonen}, {Osborne},
  {Pancino}, {Pauwels}, {Recio-Blanco}, {Reyl{\'e}}, {Riello}, {Rimoldini},
  {Roegiers}, {Rybizki}, {Sarro}, {Siopis}, {Smith}, {Sozzetti}, {Utrilla},
  {van Leeuwen}, {Abbas}, {{\'A}brah{\'a}m}, {Abreu Aramburu}, {Aerts},
  {Aguado}, {Ajaj}, {Aldea-Montero}, {Altavilla}, {{\'A}lvarez}, {Alves},
  {Anders}, {Anderson}, {Anglada Varela}, {Antoja}, {Baines}, {Baker},
  {Balaguer-N{\'u}{\~n}ez}, {Balbinot}, {Balog}, {Barache}, {Barbato},
  {Barros}, {Barstow}, {Bartolom{\'e}}, {Bassilana}, {Bauchet}, {Becciani},
  {Bellazzini}, {Berihuete}, {Bernet}, {Bertone}, {Bianchi}, {Binnenfeld},
  {Blanco-Cuaresma}, {Blazere}, {Boch}, {Bombrun}, {Bossini}, {Bouquillon},
  {Bragaglia}, {Bramante}, {Breedt}, {Bressan}, {Brouillet}, {Brugaletta},
  {Bucciarelli}, {Burlacu}, {Butkevich}, {Buzzi}, {Caffau}, {Cancelliere},
  {Cantat-Gaudin}, {Carballo}, {Carlucci}, {Carnerero}, {Carrasco},
  {Casamiquela}, {Castellani}, {Castro-Ginard}, {Chaoul}, {Charlot}, {Chemin},
  {Chiaramida}, {Chiavassa}, {Chornay}, {Comoretto}, {Contursi}, {Cooper},
  {Cornez}, {Cowell}, {Crifo}, {Cropper}, {Crosta}, {Crowley}, {Dafonte},
  {Dapergolas}, {David}, {David}, {de Laverny}, {De Luise}, {De March}, {De
  Ridder}, {de Souza}, {de Torres}, {del Peloso}, {del Pozo}, {Delbo},
  {Delgado}, {Delisle}, {Demouchy}, {Dharmawardena}, {Di Matteo}, {Diakite},
  {Diener}, {Distefano}, {Dolding}, {Edvardsson}, {Enke}, {Fabre}, {Fabrizio},
  {Faigler}, {Fedorets}, {Fernique}, {Fienga}, {Figueras}, {Fournier},
  {Fouron}, {Fragkoudi}, {Gai}, {Garcia-Gutierrez}, {Garcia-Reinaldos},
  {Garc{\'\i}a-Torres}, {Garofalo}, {Gavel}, {Gavras}, {Gerlach}, {Geyer},
  {Giacobbe}, {Gilmore}, {Girona}, {Giuffrida}, {Gomel}, {Gomez},
  {Gonz{\'a}lez-N{\'u}{\~n}ez}, {Gonz{\'a}lez-Santamar{\'\i}a},
  {Gonz{\'a}lez-Vidal}, {Granvik}, {Guillout}, {Guiraud},
  {Guti{\'e}rrez-S{\'a}nchez}, {Guy}, {Hatzidimitriou}, {Hauser}, {Haywood},
  {Helmer}, {Helmi}, {Sarmiento}, {Hidalgo}, {Hilger}, {H{\l}adczuk}, {Hobbs},
  {Holland}, {Huckle}, {Jardine}, {Jasniewicz}, {Jean-Antoine Piccolo},
  {Jim{\'e}nez-Arranz}, {Jorissen}, {Juaristi Campillo}, {Julbe}, {Karbevska},
  {Kervella}, {Khanna}, {Kontizas}, {Kordopatis}, {Korn}, {K{\'o}sp{\'a}l},
  {Kostrzewa-Rutkowska}, {Kruszy{\'n}ska}, {Kun}, {Laizeau}, {Lambert},
  {Lanza}, {Lasne}, {Le Campion}, {Lebreton}, {Lebzelter}, {Leccia}, {Leclerc},
  {Lecoeur-Taibi}, {Liao}, {Licata}, {Lindstr{\o}m}, {Lister}, {Livanou},
  {Lobel}, {Lorca}, {Loup}, {Madrero Pardo}, {Magdaleno Romeo}, {Managau},
  {Mann}, {Manteiga}, {Marchant}, {Marconi}, {Marcos}, {Marcos Santos},
  {Mar{\'\i}n Pina}, {Marinoni}, {Marocco}, {Marshall}, {Martin Polo},
  {Mart{\'\i}n-Fleitas}, {Marton}, {Mary}, {Masip}, {Massari},
  {Mastrobuono-Battisti}, {Mazeh}, {McMillan}, {Messina}, {Michalik}, {Millar},
  {Mints}, {Molina}, {Molinaro}, {Moln{\'a}r}, {Monari}, {Mongui{\'o}},
  {Montegriffo}, {Montero}, {Mor}, {Mora}, {Morbidelli}, {Morel}, {Morris},
  {Muraveva}, {Murphy}, {Musella}, {Nagy}, {Noval}, {Oca{\~n}a}, {Ogden},
  {Ordenovic}, {Osinde}, {Pagani}, {Pagano}, {Palaversa}, {Palicio},
  {Pallas-Quintela}, {Panahi}, {Payne-Wardenaar}, {Pe{\~n}alosa Esteller},
  {Penttil{\"a}}, {Pichon}, {Piersimoni}, {Pineau}, {Plachy}, {Plum}, {Poggio},
  {Pr{\v{s}}a}, {Pulone}, {Racero}, {Ragaini}, {Rainer}, {Raiteri}, {Rambaux},
  {Ramos}, {Ramos-Lerate}, {Re Fiorentin}, {Regibo}, {Richards}, {Rios Diaz},
  {Ripepi}, {Riva}, {Rix}, {Rixon}, {Robichon}, {Robin}, {Robin}, {Roelens},
  {Rogues}, {Rohrbasser}, {Romero-G{\'o}mez}, {Rowell}, {Royer}, {Ruz Mieres},
  {Rybicki}, {Sadowski}, {S{\'a}ez N{\'u}{\~n}ez}, {Sagrist{\`a} Sell{\'e}s},
  {Sahlmann}, {Salguero}, {Samaras}, {Sanchez Gimenez}, {Sanna},
  {Santove{\~n}a}, {Sarasso}, {Schultheis}, {Sciacca}, {Segol}, {Segovia},
  {S{\'e}gransan}, {Semeux}, {Shahaf}, {Siddiqui}, {Siebert}, {Siltala},
  {Silvelo}, {Slezak}, {Slezak}, {Smart}, {Snaith}, {Solano}, {Solitro},
  {Souami}, {Souchay}, {Spagna}, {Spina}, {Spoto}, {Steele},
  {Steidelm{\"u}ller}, {Stephenson}, {S{\"u}veges}, {Surdej}, {Szabados},
  {Szegedi-Elek}, {Taris}, {Taylor}, {Teixeira}, {Tolomei}, {Tonello}, {Torra},
  {Torra}, {Torralba Elipe}, {Trabucchi}, {Tsounis}, {Turon}, {Ulla}, {Unger},
  {Vaillant}, {van Dillen}, {van Reeven}, {Vanel}, {Vecchiato}, {Viala},
  {Vicente}, {Voutsinas}, {Weiler}, {Wevers}, {Wyrzykowski}, {Yoldas}, {Yvard},
  {Zhao}, {Zorec}, {Zucker}, \& {Zwitter}}]{Gaia_DR3}
{Gaia Collaboration}, {Vallenari}, A., {Brown}, A.~G.~A., {et~al.} 2023, \aap,
  674, A1, \dodoi{10.1051/0004-6361/202243940}

\bibitem[{{Gao} \& {Li}(2023)}]{Gao_2023}
{Gao}, S.-J., \& {Li}, X.-D. 2023, \mnras, 525, 2605,
  \dodoi{10.1093/mnras/stad2446}

\bibitem[{{Gianninas} {et~al.}(2014){Gianninas}, {Dufour}, {Kilic}, {Brown},
  {Bergeron}, \& {Hermes}}]{Gianninas_2014}
{Gianninas}, A., {Dufour}, P., {Kilic}, M., {et~al.} 2014, \apj, 794, 35,
  \dodoi{10.1088/0004-637X/794/1/35}

\bibitem[{{Green}(2018)}]{dustmaps}
{Green}, G. 2018, The Journal of Open Source Software, 3, 695,
  \dodoi{10.21105/joss.00695}

\bibitem[{Green {et~al.}(2019)Green, Schlafly, Zucker, Speagle, \&
  Finkbeiner}]{Green_2019}
Green, G.~M., Schlafly, E., Zucker, C., Speagle, J.~S., \& Finkbeiner, D. 2019,
  The Astrophysical Journal, 887, 93, \dodoi{10.3847/1538-4357/ab5362}

\bibitem[{Green {et~al.}(2018)Green, Schlafly, Finkbeiner, Rix, Martin,
  Burgett, Draper, Flewelling, Hodapp, Kaiser, Kudritzki, Magnier, Metcalfe,
  Tonry, Wainscoat, \& Waters}]{Green_2018}
Green, G.~M., Schlafly, E.~F., Finkbeiner, D., {et~al.} 2018, Monthly Notices
  of the Royal Astronomical Society, 478, 651, \dodoi{10.1093/mnras/sty1008}

\bibitem[{{Green} {et~al.}(2023){Green}, {Maoz}, {Mazeh}, {Faigler}, {Shahaf},
  {Gomel}, {El-Badry}, \& {Rix}}]{green_2023}
{Green}, M.~J., {Maoz}, D., {Mazeh}, T., {et~al.} 2023, \mnras, 522, 29,
  \dodoi{10.1093/mnras/stad915}

\bibitem[{{Gunn} {et~al.}(2006){Gunn}, {Siegmund}, {Mannery}, {Owen}, {Hull},
  {Leger}, {Carey}, {Knapp}, {York}, {Boroski}, {Kent}, {Lupton}, {Rockosi},
  {Evans}, {Waddell}, {Anderson}, {Annis}, {Barentine}, {Bartoszek}, {Bastian},
  {Bracker}, {Brewington}, {Briegel}, {Brinkmann}, {Brown}, {Carr},
  {Czarapata}, {Drennan}, {Dombeck}, {Federwitz}, {Gillespie}, {Gonzales},
  {Hansen}, {Harvanek}, {Hayes}, {Jordan}, {Kinney}, {Klaene}, {Kleinman},
  {Kron}, {Kresinski}, {Lee}, {Limmongkol}, {Lindenmeyer}, {Long}, {Loomis},
  {McGehee}, {Mantsch}, {Neilsen}, {Neswold}, {Newman}, {Nitta}, {Peoples},
  {Pier}, {Prieto}, {Prosapio}, {Rivetta}, {Schneider}, {Snedden}, \&
  {Wang}}]{gunn_2006}
{Gunn}, J.~E., {Siegmund}, W.~A., {Mannery}, E.~J., {et~al.} 2006, \aj, 131,
  2332, \dodoi{10.1086/500975}

\bibitem[{Harris {et~al.}(2020)Harris, Millman, van~der Walt, Gommers,
  Virtanen, Cournapeau, Wieser, Taylor, Berg, Smith, Kern, Picus, Hoyer, van
  Kerkwijk, Brett, Haldane, del R{\'{i}}o, Wiebe, Peterson,
  G{\'{e}}rard-Marchant, Sheppard, Reddy, Weckesser, Abbasi, Gohlke, \&
  Oliphant}]{numpy}
Harris, C.~R., Millman, K.~J., van~der Walt, S.~J., {et~al.} 2020, Nature, 585,
  357, \dodoi{10.1038/s41586-020-2649-2}

\bibitem[{{Heber}(2009)}]{Heber_2009}
{Heber}, U. 2009, \araa, 47, 211, \dodoi{10.1146/annurev-astro-082708-101836}

\bibitem[{{Hilditch} {et~al.}(1996){Hilditch}, {Harries}, \&
  {Hill}}]{Hilditch_1996}
{Hilditch}, R.~W., {Harries}, T.~J., \& {Hill}, G. 1996, \mnras, 279, 1380,
  \dodoi{10.1093/mnras/279.4.1380}

\bibitem[{Hunter(2007)}]{matplotlib}
Hunter, J.~D. 2007, Computing in Science \& Engineering, 9, 90,
  \dodoi{10.1109/MCSE.2007.55}

\bibitem[{{Iben} \& {Tutukov}(1985)}]{Iben_1985}
{Iben}, I., J., \& {Tutukov}, A.~V. 1985, \apjs, 58, 661,
  \dodoi{10.1086/191054}

\bibitem[{{IRSA}(2022)}]{ZTF_doi}
{IRSA}. 2022, Zwicky Transient Facility Image Service,  IPAC,
  \dodoi{10.26131/IRSA539}

\bibitem[{Istrate {et~al.}(2014{\natexlab{a}})Istrate, M~Tauris, \&
  Langer}]{istrate_formation_2014}
Istrate, A.~G., M~Tauris, T., \& Langer, N. 2014{\natexlab{a}}, A\&A, 571, A45,
  \dodoi{10.1051/0004-6361/201424680}

\bibitem[{Istrate {et~al.}(2016)Istrate, Marchant, Tauris, Langer, Stancliffe,
  \& Grassitelli}]{Istrate_2016}
Istrate, A.~G., Marchant, P., Tauris, T.~M., {et~al.} 2016, Astronomy {\&}
  Astrophysics, 595, A35, \dodoi{10.1051/0004-6361/201628874}

\bibitem[{Istrate {et~al.}(2014{\natexlab{b}})Istrate, Tauris, Langer, \&
  Antoniadis}]{istrate_timescale_2014}
Istrate, A.~G., Tauris, T.~M., Langer, N., \& Antoniadis, J.
  2014{\natexlab{b}}, A\&A, 571, L3, \dodoi{10.1051/0004-6361/201424681}

\bibitem[{{Jarrett} {et~al.}(2011){Jarrett}, {Cohen}, {Masci}, {Wright},
  {Stern}, {Benford}, {Blain}, {Carey}, {Cutri}, {Eisenhardt}, {Lonsdale},
  {Mainzer}, {Marsh}, {Padgett}, {Petty}, {Ressler}, {Skrutskie}, {Stanford},
  {Surace}, {Tsai}, {Wheelock}, \& {Yan}}]{WISE_2011}
{Jarrett}, T.~H., {Cohen}, M., {Masci}, F., {et~al.} 2011, \apj, 735, 112,
  \dodoi{10.1088/0004-637X/735/2/112}

\bibitem[{{Jha} {et~al.}(2019){Jha}, {Maguire}, \& {Sullivan}}]{Jha_2019}
{Jha}, S.~W., {Maguire}, K., \& {Sullivan}, M. 2019, Nature Astronomy, 3, 706,
  \dodoi{10.1038/s41550-019-0858-0}

\bibitem[{{Kaltenborn} {et~al.}(2022){Kaltenborn}, {Fryer}, {Wollaeger},
  {Belczynski}, {Even}, \& {Kouveliotou}}]{Kaltenbourn_2022}
{Kaltenborn}, M.~A., {Fryer}, C.~L., {Wollaeger}, R.~T., {et~al.} 2022, arXiv
  e-prints, arXiv:2209.13061, \dodoi{10.48550/arXiv.2209.13061}

\bibitem[{Kanodia \& Wright(2018)}]{Kanodia_2018}
Kanodia, S., \& Wright, J. 2018, Research Notes of the AAS, 2, 4,
  \dodoi{10.3847/2515-5172/aaa4b7}

\bibitem[{Kilic {et~al.}(2011)Kilic, Brown, Prieto, Agüeros, Heinke, \&
  Kenyon}]{kilic_elm_2011}
Kilic, M., Brown, W.~R., Prieto, C.~A., {et~al.} 2011, ApJ, 727, 3,
  \dodoi{10.1088/0004-637X/727/1/3}

\bibitem[{{Koester}(2010)}]{Koester_2010}
{Koester}, D. 2010, \memsai, 81, 921

\bibitem[{{Kollmeier} {et~al.}(2019){Kollmeier}, {Anderson}, {Blanc},
  {Blanton}, {Covey}, {Crane}, {Drory}, {Frinchaboy}, {Froning}, {Johnson},
  {Kneib}, {Kreckel}, {Merloni}, {Pellegrini}, {Pogge}, {Ramirez}, {Rix},
  {Sayres}, {S{\'a}nchez-Gallego}, {Shen}, {Tkachenko}, {Trump}, {Tuttle},
  {Weijmans}, {Zasowski}, {Barbuy}, {Beaton}, {Bergemann}, {Bochanski},
  {Brandt}, {Casey}, {Cherinka}, {Eracleous}, {Fan}, {Garc{\'\i}a}, {Green},
  {Hekker}, {Lane}, {Longa-Pe{\~n}a}, {Mathur}, {Meza}, {Minchev}, {Myers},
  {Nidever}, {Nitschelm}, {O'Connell}, {Price-Whelan}, {Raddick}, {Rossi},
  {Sankrit}, {Simon}, {Stutz}, {Ting}, {Trakhtenbrot}, {Weaver}, {Willmer}, \&
  {Weinberg}}]{Kollmeier_2019}
{Kollmeier}, J., {Anderson}, S.~F., {Blanc}, G.~A., {et~al.} 2019, in Bulletin
  of the American Astronomical Society, Vol.~51, 274

\bibitem[{Kollmeier {et~al.}(2017)Kollmeier, Zasowski, Rix, Johns, Anderson,
  Drory, Johnson, Pogge, Bird, Blanc, Brownstein, Crane, De~Lee, Klaene,
  Kreckel, MacDonald, Merloni, Ness, O'Brien, Sanchez-Gallego, Sayres, Shen,
  Thakar, Tkachenko, Aerts, Blanton, Eisenstein, Holtzman, Maoz, Nandra,
  Rockosi, Weinberg, Bovy, Casey, Chaname, Clerc, Conroy, Eracleous, Gänsicke,
  Hekker, Horne, Kauffmann, McQuinn, Pellegrini, Schinnerer, Schlafly, Schwope,
  Seibert, Teske, \& van Saders}]{kollmeier_sdss-v_2017}
Kollmeier, J.~A., Zasowski, G., Rix, H.-W., {et~al.} 2017, {SDSS}-{V}:
  {Pioneering} {Panoptic} {Spectroscopy},  arXiv.
\newblock \url{http://arxiv.org/abs/1711.03234}

\bibitem[{{Kopal}(1959)}]{kopal_1959}
{Kopal}, Z. 1959, {Close binary systems}

\bibitem[{{Korol} {et~al.}(2022){Korol}, {Hallakoun}, {Toonen}, \&
  {Karnesis}}]{Korol_2022}
{Korol}, V., {Hallakoun}, N., {Toonen}, S., \& {Karnesis}, N. 2022, \mnras,
  511, 5936, \dodoi{10.1093/mnras/stac415}

\bibitem[{{Kupfer} {et~al.}(2020{\natexlab{a}}){Kupfer}, {Bauer}, {Marsh}, {van
  Roestel}, {Bellm}, {Burdge}, {Coughlin}, {Fuller}, {Hermes}, {Bildsten},
  {Kulkarni}, {Prince}, {Szkody}, {Dhillon}, {Murawski}, {Burruss}, {Dekany},
  {Delacroix}, {Drake}, {Duev}, {Feeney}, {Graham}, {Kaplan}, {Laher},
  {Littlefair}, {Masci}, {Riddle}, {Rusholme}, {Serabyn}, {Smith}, {Shupe}, \&
  {Soumagnac}}]{Kupfer_2020a}
{Kupfer}, T., {Bauer}, E.~B., {Marsh}, T.~R., {et~al.} 2020{\natexlab{a}},
  \apj, 891, 45, \dodoi{10.3847/1538-4357/ab72ff}

\bibitem[{{Kupfer} {et~al.}(2020{\natexlab{b}}){Kupfer}, {Bauer}, {Burdge},
  {Roestel}, {Bellm}, {Fuller}, {Hermes}, {Marsh}, {Bildsten}, {Kulkarni},
  {Phinney}, {Prince}, {Szkody}, {Yao}, {Irrgang}, {Heber}, {Schneider},
  {Dhillon}, {Murawski}, {Drake}, {Duev}, {Feeney}, {Graham}, {Laher},
  {Littlefair}, {Mahabal}, {Masci}, {Porter}, {Reiley}, {Rodriguez},
  {Rusholme}, {Shupe}, \& {Soumagnac}}]{Kupfer_2020b}
{Kupfer}, T., {Bauer}, E.~B., {Burdge}, K.~B., {et~al.} 2020{\natexlab{b}},
  \apjl, 898, L25, \dodoi{10.3847/2041-8213/aba3c2}

\bibitem[{{Lagos} {et~al.}(2020){Lagos}, {Schreiber}, {Parsons},
  {G{\"a}nsicke}, \& {Godoy}}]{Lagos_2020}
{Lagos}, F., {Schreiber}, M.~R., {Parsons}, S.~G., {G{\"a}nsicke}, B.~T., \&
  {Godoy}, N. 2020, \mnras, 499, L121, \dodoi{10.1093/mnrasl/slaa164}

\bibitem[{{Lallement} {et~al.}(2019){Lallement}, {Babusiaux}, {Vergely},
  {Katz}, {Arenou}, {Valette}, {Hottier}, \& {Capitanio}}]{Lallement_2019}
{Lallement}, R., {Babusiaux}, C., {Vergely}, J.~L., {et~al.} 2019, \aap, 625,
  A135, \dodoi{10.1051/0004-6361/201834695}

\bibitem[{{Lee} {et~al.}(2018){Lee}, {Hui}, {Takata}, {Kong}, {Tam}, \&
  {Cheng}}]{Lee_2018}
{Lee}, J., {Hui}, C.~Y., {Takata}, J., {et~al.} 2018, \apj, 864, 23,
  \dodoi{10.3847/1538-4357/aad284}

\bibitem[{Lejeune {et~al.}(1997)Lejeune, Cuisinier, \& Buser}]{Lejeune_1997}
Lejeune, T., Cuisinier, F., \& Buser, R. 1997, Astron. Astrophys. Suppl. Ser.,
  125, 229, \dodoi{10.1051/aas:1997373}

\bibitem[{Lejeune {et~al.}(1998)Lejeune, Cuisinier, \& Buser}]{Lejeune_1998}
---. 1998, Astron. Astrophys. Suppl. Ser., 130, 65, \dodoi{10.1051/aas:1998405}

\bibitem[{Li {et~al.}(2019)Li, Chen, Chen, \& Han}]{li_formation_2019}
Li, Z., Chen, X., Chen, H.-L., \& Han, Z. 2019, ApJ, 871, 148,
  \dodoi{10.3847/1538-4357/aaf9a1}

\bibitem[{{Lightkurve Collaboration} {et~al.}(2018){Lightkurve Collaboration},
  {Cardoso}, {Hedges}, {Gully-Santiago}, {Saunders}, {Cody}, {Barclay}, {Hall},
  {Sagear}, {Turtelboom}, {Zhang}, {Tzanidakis}, {Mighell}, {Coughlin}, {Bell},
  {Berta-Thompson}, {Williams}, {Dotson}, \& {Barentsen}}]{lightkurve}
{Lightkurve Collaboration}, {Cardoso}, J.~V.~d.~M., {Hedges}, C., {et~al.}
  2018, {Lightkurve: Kepler and TESS time series analysis in Python},
  Astrophysics Source Code Library.
\newblock \doeprint{1812.013}

\bibitem[{{Liu} \& {Chen}(2011)}]{Liu_2011}
{Liu}, W.-M., \& {Chen}, W.-C. 2011, \mnras, 416, 2285,
  \dodoi{10.1111/j.1365-2966.2011.19202.x}

\bibitem[{Maoz {et~al.}(2014)Maoz, Mannucci, \&
  Nelemans}]{maoz_observational_2014}
Maoz, D., Mannucci, F., \& Nelemans, G. 2014, Annu. Rev. Astron. Astrophys.,
  52, 107, \dodoi{10.1146/annurev-astro-082812-141031}

\bibitem[{{Margalit} \& {Metzger}(2016)}]{Margalit_2016}
{Margalit}, B., \& {Metzger}, B.~D. 2016, \mnras, 461, 1154,
  \dodoi{10.1093/mnras/stw1410}

\bibitem[{{Marsh}(2001)}]{Marsh_2001}
{Marsh}, T.~R. 2001, \mnras, 324, 547, \dodoi{10.1046/j.1365-8711.2001.04293.x}

\bibitem[{{Marsh}(2011)}]{Marsh_2011}
---. 2011, Classical and Quantum Gravity, 28, 094019,
  \dodoi{10.1088/0264-9381/28/9/094019}

\bibitem[{{Marsh} {et~al.}(1995){Marsh}, {Dhillon}, \& {Duck}}]{Marsh_1995}
{Marsh}, T.~R., {Dhillon}, V.~S., \& {Duck}, S.~R. 1995, \mnras, 275, 828,
  \dodoi{10.1093/mnras/275.3.828}

\bibitem[{Masci {et~al.}(2019)Masci, Laher, Rusholme, Shupe, Groom, Surace,
  Jackson, Monkewitz, Beck, Flynn, Terek, Landry, Hacopians, Desai, Howell,
  Brooke, Imel, Wachter, Ye, Lin, Cenko, Cunningham, Rebbapragada, Bue, Miller,
  Mahabal, Bellm, Patterson, Jurić, Golkhou, Ofek, Walters, Graham, Kasliwal,
  Dekany, Kupfer, Burdge, Cannella, Barlow, Sistine, Giomi, Fremling,
  Blagorodnova, Levitan, Riddle, Smith, Helou, Prince, \&
  Kulkarni}]{masci_zwicky_2019}
Masci, F.~J., Laher, R.~R., Rusholme, B., {et~al.} 2019, PASP, 131, 018003,
  \dodoi{10.1088/1538-3873/aae8ac}

\bibitem[{{Maxted} {et~al.}(2011){Maxted}, {Anderson}, {Burleigh}, {Collier
  Cameron}, {Heber}, {G{\"a}nsicke}, {Geier}, {Kupfer}, {Marsh}, {Nelemans},
  {O'Toole}, {{\O}stensen}, {Smalley}, \& {West}}]{Maxted_2011}
{Maxted}, P.~F.~L., {Anderson}, D.~R., {Burleigh}, M.~R., {et~al.} 2011,
  \mnras, 418, 1156, \dodoi{10.1111/j.1365-2966.2011.19567.x}

\bibitem[{{Nelson} {et~al.}(2004){Nelson}, {Dubeau}, \&
  {MacCannell}}]{Nelson_2004}
{Nelson}, L.~A., {Dubeau}, E., \& {MacCannell}, K.~A. 2004, \apj, 616, 1124,
  \dodoi{10.1086/421698}

\bibitem[{{Nomoto}(1984)}]{Nomoto_1984}
{Nomoto}, K. 1984, \apj, 277, 791, \dodoi{10.1086/161749}

\bibitem[{Ochsenbein {et~al.}(2000)Ochsenbein, Bauer, \&
  Marcout}]{Ochsenbein_2000}
Ochsenbein, F., Bauer, P., \& Marcout, J. 2000, Astronomy and Astrophysics
  Supplement Series, 143, 23, \dodoi{10.1051/aas:2000169}

\bibitem[{Poznanski {et~al.}(2012)Poznanski, Prochaska, \&
  Bloom}]{Poznanski_2012}
Poznanski, D., Prochaska, J.~X., \& Bloom, J.~S. 2012, Monthly Notices of the
  Royal Astronomical Society, 426, 1465,
  \dodoi{10.1111/j.1365-2966.2012.21796.x}

\bibitem[{Prša(2018)}]{prsa_modeling_2018}
Prša, A. 2018, Modeling and Analysis of Eclipsing Binary Stars, 2514-3433 (IOP
  Publishing), \dodoi{10.1088/978-0-7503-1287-5}

\bibitem[{{Ricker} {et~al.}(2015){Ricker}, {Winn}, {Vanderspek}, {Latham},
  {Bakos}, {Bean}, {Berta-Thompson}, {Brown}, {Buchhave}, {Butler}, {Butler},
  {Chaplin}, {Charbonneau}, {Christensen-Dalsgaard}, {Clampin}, {Deming},
  {Doty}, {De Lee}, {Dressing}, {Dunham}, {Endl}, {Fressin}, {Ge}, {Henning},
  {Holman}, {Howard}, {Ida}, {Jenkins}, {Jernigan}, {Johnson}, {Kaltenegger},
  {Kawai}, {Kjeldsen}, {Laughlin}, {Levine}, {Lin}, {Lissauer}, {MacQueen},
  {Marcy}, {McCullough}, {Morton}, {Narita}, {Paegert}, {Palle}, {Pepe},
  {Pepper}, {Quirrenbach}, {Rinehart}, {Sasselov}, {Sato}, {Seager},
  {Sozzetti}, {Stassun}, {Sullivan}, {Szentgyorgyi}, {Torres}, {Udry}, \&
  {Villasenor}}]{TESS_2015}
{Ricker}, G.~R., {Winn}, J.~N., {Vanderspek}, R., {et~al.} 2015, Journal of
  Astronomical Telescopes, Instruments, and Systems, 1, 014003,
  \dodoi{10.1117/1.JATIS.1.1.014003}

\bibitem[{{Scargle}(1982)}]{Scargle_1982}
{Scargle}, J.~D. 1982, \apj, 263, 835, \dodoi{10.1086/160554}

\bibitem[{{Schaffenroth} {et~al.}(2022){Schaffenroth}, {Pelisoli}, {Barlow},
  {Geier}, \& {Kupfer}}]{Schaffenroth_2022}
{Schaffenroth}, V., {Pelisoli}, I., {Barlow}, B.~N., {Geier}, S., \& {Kupfer},
  T. 2022, \aap, 666, A182, \dodoi{10.1051/0004-6361/202244214}

\bibitem[{{Shao} \& {Li}(2012)}]{Shao_2012}
{Shao}, Y., \& {Li}, X.-D. 2012, \apj, 756, 85,
  \dodoi{10.1088/0004-637X/756/1/85}

\bibitem[{{Skrutskie} {et~al.}(2006){Skrutskie}, {Cutri}, {Stiening},
  {Weinberg}, {Schneider}, {Carpenter}, {Beichman}, {Capps}, {Chester},
  {Elias}, {Huchra}, {Liebert}, {Lonsdale}, {Monet}, {Price}, {Seitzer},
  {Jarrett}, {Kirkpatrick}, {Gizis}, {Howard}, {Evans}, {Fowler}, {Fullmer},
  {Hurt}, {Light}, {Kopan}, {Marsh}, {McCallon}, {Tam}, {Van Dyk}, \&
  {Wheelock}}]{2MASS_2006}
{Skrutskie}, M.~F., {Cutri}, R.~M., {Stiening}, R., {et~al.} 2006, \aj, 131,
  1163, \dodoi{10.1086/498708}

\bibitem[{Smee {et~al.}(2013)Smee, Gunn, Uomoto, Roe, Schlegel, Rockosi, Carr,
  Leger, Dawson, Olmstead, Brinkmann, Owen, Barkhouser, Honscheid, Harding,
  Long, Lupton, Loomis, Anderson, Annis, Bernardi, Bhardwaj, Bizyaev, Bolton,
  Brewington, Briggs, Burles, Burns, Castander, Connolly, Davenport, Ebelke,
  Epps, Feldman, Friedman, Frieman, Heckman, Hull, Knapp, Lawrence, Loveday,
  Mannery, Malanushenko, Malanushenko, Merrelli, Muna, Newman, Nichol, Oravetz,
  Pan, Pope, Ricketts, Shelden, Sandford, Siegmund, Simmons, Smith, Snedden,
  Schneider, SubbaRao, Tremonti, Waddell, \& York}]{Smee_2013}
Smee, S.~A., Gunn, J.~E., Uomoto, A., {et~al.} 2013, The Astronomical Journal,
  146, 32, \dodoi{10.1088/0004-6256/146/2/32}

\bibitem[{{Soker}(2019)}]{Soker_2019}
{Soker}, N. 2019, \nar, 87, 101535, \dodoi{10.1016/j.newar.2020.101535}

\bibitem[{{STScI}(2019)}]{TESS_18_DOI}
{STScI}. 2019, TESS Calibrated Full Frame Images: Sector 18,  STScI/MAST,
  \dodoi{10.17909/2H0M-QR07}

\bibitem[{{STScI}(2020)}]{TESS_19_DOI}
---. 2020, TESS Calibrated Full Frame Images: Sector 19,  STScI/MAST,
  \dodoi{10.17909/MSXY-D755}

\bibitem[{{Truemper}(1982)}]{Truemper_1982}
{Truemper}, J. 1982, Advances in Space Research, 2, 241,
  \dodoi{10.1016/0273-1177(82)90070-9}

\bibitem[{{VanderPlas}(2018)}]{VanderPlas_2018}
{VanderPlas}, J.~T. 2018, \apjs, 236, 16, \dodoi{10.3847/1538-4365/aab766}

\bibitem[{{Vaz}(1985)}]{Vaz_1985}
{Vaz}, L.~P.~R. 1985, \apss, 113, 349, \dodoi{10.1007/BF00650970}

\bibitem[{Virtanen {et~al.}(2020{\natexlab{a}})Virtanen, Gommers, Oliphant,
  Haberland, Reddy, Cournapeau, Burovski, Peterson, Weckesser, Bright, {van der
  Walt}, Brett, Wilson, Millman, Mayorov, Nelson, Jones, Kern, Larson, Carey,
  Polat, Feng, Moore, {VanderPlas}, Laxalde, Perktold, Cimrman, Henriksen,
  Quintero, Harris, Archibald, Ribeiro, Pedregosa, {van Mulbregt}, \& {SciPy
  1.0 Contributors}}]{2020SciPy-NMeth}
Virtanen, P., Gommers, R., Oliphant, T.~E., {et~al.} 2020{\natexlab{a}}, Nature
  Methods, 17, 261, \dodoi{10.1038/s41592-019-0686-2}

\bibitem[{Virtanen {et~al.}(2020{\natexlab{b}})Virtanen, Gommers, Oliphant,
  Haberland, Reddy, Cournapeau, Burovski, Peterson, Weckesser, Bright, {van der
  Walt}, Brett, Wilson, Millman, Mayorov, Nelson, Jones, Kern, Larson, Carey,
  Polat, Feng, Moore, {VanderPlas}, Laxalde, Perktold, Cimrman, Henriksen,
  Quintero, Harris, Archibald, Ribeiro, Pedregosa, {van Mulbregt}, \& {SciPy
  1.0 Contributors}}]{scipy}
---. 2020{\natexlab{b}}, Nature Methods, 17, 261,
  \dodoi{10.1038/s41592-019-0686-2}

\bibitem[{{von Zeipel}(1924)}]{Zeipel_1924}
{von Zeipel}, H. 1924, \mnras, 84, 665, \dodoi{10.1093/mnras/84.9.665}

\bibitem[{{Wang} {et~al.}(2020){Wang}, {Gies}, {Lester}, {Guo}, {Matson},
  {Peters}, {Dhillon}, {Butterley}, {Littlefair}, {Wilson}, \&
  {Maxted}}]{Wang_2020}
{Wang}, L., {Gies}, D.~R., {Lester}, K.~V., {et~al.} 2020, \aj, 159, 4,
  \dodoi{10.3847/1538-3881/ab52fa}

\bibitem[{Weiser \& Zarantonello(1988)}]{Weiser_1998}
Weiser, A., \& Zarantonello, S.~E. 1988, Mathematics of Computation, 50, 189.
\newblock \url{http://www.jstor.org/stable/2007922}

\bibitem[{{Wilson} {et~al.}(2019){Wilson}, {Hearty}, {Skrutskie}, {Majewski},
  {Holtzman}, {Eisenstein}, {Gunn}, {Blank}, {Henderson}, {Smee}, {Nelson},
  {Nidever}, {Arns}, {Barkhouser}, {Barr}, {Beland}, {Bershady}, {Blanton},
  {Brunner}, {Burton}, {Carey}, {Carr}, {Colque}, {Crane}, {Damke}, {Davidson},
  {Dean}, {Di Mille}, {Don}, {Ebelke}, {Evans}, {Fitzgerald}, {Gillespie},
  {Hall}, {Harding}, {Harding}, {Hammond}, {Hancock}, {Harrison}, {Hope},
  {Horne}, {Karakla}, {Lam}, {Leger}, {MacDonald}, {Maseman}, {Matsunari},
  {Melton}, {Mitcheltree}, {O'Brien}, {O'Connell}, {Patten}, {Richardson},
  {Rieke}, {Rieke}, {Roman-Lopes}, {Schiavon}, {Sobeck}, {Stolberg}, {Stoll},
  {Tembe}, {Trujillo}, {Uomoto}, {Vernieri}, {Walker}, {Weinberg}, {Young},
  {Anthony-Brumfield}, {Bizyaev}, {Breslauer}, {De Lee}, {Downey}, {Halverson},
  {Huehnerhoff}, {Klaene}, {Leon}, {Long}, {Mahadevan}, {Malanushenko},
  {Nguyen}, {Owen}, {S{\'a}nchez-Gallego}, {Sayres}, {Shane}, {Shectman},
  {Shetrone}, {Skinner}, {Stauffer}, \& {Zhao}}]{wilson_2019}
{Wilson}, J.~C., {Hearty}, F.~R., {Skrutskie}, M.~F., {et~al.} 2019, \pasp,
  131, 055001, \dodoi{10.1088/1538-3873/ab0075}

\bibitem[{{Wilson}(1990)}]{Wilson_1990}
{Wilson}, R.~E. 1990, \apj, 356, 613, \dodoi{10.1086/168867}

\bibitem[{{Wright} {et~al.}(2010){Wright}, {Eisenhardt}, {Mainzer}, {Ressler},
  {Cutri}, {Jarrett}, {Kirkpatrick}, {Padgett}, {McMillan}, {Skrutskie},
  {Stanford}, {Cohen}, {Walker}, {Mather}, {Leisawitz}, {Gautier}, {McLean},
  {Benford}, {Lonsdale}, {Blain}, {Mendez}, {Irace}, {Duval}, {Liu}, {Royer},
  {Heinrichsen}, {Howard}, {Shannon}, {Kendall}, {Walsh}, {Larsen}, {Cardon},
  {Schick}, {Schwalm}, {Abid}, {Fabinsky}, {Naes}, \& {Tsai}}]{Wright_2010}
{Wright}, E.~L., {Eisenhardt}, P. R.~M., {Mainzer}, A.~K., {et~al.} 2010, \aj,
  140, 1868, \dodoi{10.1088/0004-6256/140/6/1868}

\bibitem[{{Yamaguchi} {et~al.}(2023){Yamaguchi}, {El-Badry}, {Rodriguez},
  {Gull}, {Roulston}, \& {Vanderbosch}}]{Yamaguchi_2023}
{Yamaguchi}, N., {El-Badry}, K., {Rodriguez}, A.~C., {et~al.} 2023, \mnras,
  524, 740, \dodoi{10.1093/mnras/stad1878}

\bibitem[{{Zavlin}(2006)}]{Zavlin_2006}
{Zavlin}, V.~E. 2006, \apj, 638, 951, \dodoi{10.1086/449308}

\bibitem[{{Zavlin} {et~al.}(2002){Zavlin}, {Pavlov}, {Sanwal}, {Manchester},
  {Tr{\"u}mper}, {Halpern}, \& {Becker}}]{Zavlin_2002}
{Zavlin}, V.~E., {Pavlov}, G.~G., {Sanwal}, D., {et~al.} 2002, \apj, 569, 894,
  \dodoi{10.1086/339351}

\bibitem[{{Zenati} {et~al.}(2020){Zenati}, {Bobrick}, \&
  {Perets}}]{Zenati_2020}
{Zenati}, Y., {Bobrick}, A., \& {Perets}, H.~B. 2020, \mnras, 493, 3956,
  \dodoi{10.1093/mnras/staa507}

\bibitem[{{Zenati} {et~al.}(2019){Zenati}, {Perets}, \& {Toonen}}]{Zenati_2019}
{Zenati}, Y., {Perets}, H.~B., \& {Toonen}, S. 2019, \mnras, 486, 1805,
  \dodoi{10.1093/mnras/stz316}

\end{thebibliography}
\bibliographystyle{aasjournal}



\end{document}